\documentclass[a4paper,fleqn,usenatbib,useAMS]{mnras}
\usepackage{graphicx}	
\usepackage{amsmath}	
\usepackage{amssymb}	
\usepackage{multicol}        
\usepackage{bm}		
\usepackage{pdflscape}	
\usepackage{array}
\usepackage{fancyvrb}
\usepackage{epstopdf}
\usepackage{url}
\usepackage{hyperref}
\usepackage[T1]{fontenc}
\usepackage{ae,aecompl}
\usepackage{txfonts}
\usepackage{threeparttable}
\usepackage{breakurl}
\usepackage{multirow}
\usepackage{subfig}
\usepackage{xcolor}

\newcommand{\teff}{$T_{\rm{eff}}$}
\newcommand{\logg}{$\rm{\log g}$}
\newcommand{\feh}{[Fe/H]}
\newcommand{\alfe}{\rm{[$\alpha$/Fe]}}

\title[Stellar Spectral Interpolation]
{Stellar Spectral Interpolation using Machine Learning}
\author[Sharma et al.]{Kaushal Sharma$^{1}$,\thanks{E-mail:kaushals@iucaa.in}
Harinder P. Singh$^{2}$,
Ranjan Gupta$^{1}$,
Ajit Kembhavi${^1}$,
\newauthor Kaustubh Vaghmare${^{1,3}}$,
Jianrong Shi${^4}$, 
Yongheng Zhao${^4}$,
Jiannan Zhang${^4}$, and
\newauthor Yue Wu$^4$\\
\\
$^{1}$Inter University Centre for Astronomy and Astrophysics, Pune 411007, India\\
$^{2}$Department of Physics and Astrophysics, University of Delhi, Delhi 110007, India\\
$^{3}$Persistent Systems Ltd., Pune, India\\
$^{4}$National Astronomical Observatories, Chinese Academy of Sciences, Beijing 100012, China
}

\begin{document}

\date{Received on \today; Accepted on }

\pagerange{\pageref{firstpage}--\pageref{lastpage}} \pubyear{2020}

\maketitle

\label{firstpage}

\begin{abstract}

Theoretical stellar spectra rely on model stellar atmospheres computed based on our understanding of the physical laws at play in the stellar interiors. These models, coupled with atomic and molecular line databases, are used to generate theoretical stellar spectral libraries (SSLs) comprising of stellar spectra over a regular grid of atmospheric parameters (temperature, surface gravity, abundances) at any desired resolution. Another class of SSLs is referred to as empirical spectral libraries; these contain observed spectra at limited resolution. SSLs play an essential role in deriving the properties of stars and stellar populations. Both theoretical and empirical libraries suffer from limited coverage over the parameter space. This limitation is overcome to some extent by generating spectra for specific sets of atmospheric parameters by interpolating within the grid of available parameter space.
In this work, we present a method for spectral interpolation in the optical region using machine learning algorithms that are generic, easily adaptable for any SSL without much change in the model parameters, and computationally inexpensive. We use two machine learning techniques, Random Forest (RF) and Artificial Neural Networks (ANN), and train the models on the MILES library. We apply the trained models to spectra from the CFLIB for testing and show that the performance of the two models is comparable. We show that both the models achieve better accuracy than the existing methods of polynomial based interpolation and the Gaussian radial basis function (RBF) interpolation.

\end{abstract}


\section{Introduction}\label{sec:introduction}

Empirical and synthetic stellar spectral libraries (SSLs) consisting of stellar spectra with their atmospheric parameters and/or spectral classes have played a key role in exploring the physics of stars and stellar populations through various studies. Empirical libraries such as the Indo-U.S. Library of Coud{\'e} Feed Stellar Spectra \citep[CFLIB;][]{Valdes2004}, MILES \citep{miles2006}, and ELODIE \citep{Prugniel2007} comprise of observed stellar spectra and span a wide range of parameter space in the H-R diagram. Empirical SSLs have been extensively used in the past for stellar classification \citep{Gulati1994,Bailer-Jones1998,Singh1998,Navarro2012,Liu2015}, parameter determination \citep{Wu2011,Prugniel2011,Sharma2016}, chemical and evolutionary studies of stellar populations and galaxies \citep{Buzzoni1994,Koleva2008}. Some studies involving stellar populations require the presence of stellar spectra at the edges of the parameter space (e.g., for brown dwarfs) or specific wavelength coverage in high-resolution for studying detailed abundances. In such scenarios, synthetic libraries are found to be more useful. Primary ingredients for developing synthetic SSLs are models of stellar atmosphere, e.g. ATLAS \citep{Atlas1993}, PHOENIX \citep{Hauschildt1999}, MARCS \citep{Marcs2008}, which are used by computer programs to generate synthetic stellar spectra to create synthetic spectral libraries \citep{Munari2005,Husser2013,Coelho2014}.
Synthetic libraries do not have the limitation of spectral resolution or wavelength coverage. They can provide the spectrum in the regions of the parameter space (say extreme edges of metallicity bins) which are sparsely populated in the empirical libraries. Therefore, the synthetic libraries can be used for e.g., the study of young, metal-poor stellar populations, which are rare in empirical libraries \citep{Crowl2008}. Synthetic libraries have also been used for building stellar population models \citep{Leitherer1999,Percival2009}

The presence of real features in empirical spectra and the absence of assumptions about the stellar photosphere give an edge to empirical libraries over the synthetic SSLs. However, the parameter space (\teff, \logg, and \feh{}) is non-uniformly covered by these libraries and spectra in a region of the parameter space might not be enough to conduct a specific type of study. The problem of obtaining a real spectrum for any given set of atmospheric parameters (\teff, \logg, \feh) can be resolved by the idea of spectral interpolation over the three-dimensional grid of atmospheric parameters, proposed in \citet{Koleva2009,Prugniel2011}. Such a spectrum will contain real features and can be further used for parameter determination as well as for classification. Spectral interpolation is done using a polynomial \citep[See Eq. 3,][]{Prugniel2011} as a function of \teff, \logg, \feh{} and $\lambda$. Development of such a polynomial based interpolator, referred to as TGM in \citet{Prugniel2011}, is a cumbersome process and requires fine-tuning/addition of the terms for a specific type of stars or at the edges of the parameter space. A revised version of the interpolator, TGM2, was presented in \citet{Sharma2016} which is an improved version for cool stars considering their importance for galactic \citep{Chabrier2003,West2006} and exoplanetary studies \citep{Bonfils2005,Neves2013}. 

In this paper, we explore spectral interpolation using machine-learning (ML) algorithms. With the ML approach, one need not develop a polynomial for mapping the three atmospheric parameters (\teff, \logg, \feh) to the flux values. Instead, the transformation takes place through mapping rules automatically captured and learned from the dataset by the ML algorithms. Machine learning based approaches are easily generalizable and can be implemented for any SSL to generate an interpolated spectrum for a given set of parameters. 

ML approaches are generally put into three broad categories: supervised learning, unsupervised learning, and reinforcement learning. The  supervised algorithms involve training on a `labelled' dataset to perform a classification or regression task. Training involves optimizing the free parameters or `weights' of the model using the input data. The `labelled' input data here implies that corresponding to each input sample described by a set of features, there exists an output class for classification problems or continuous variable(s) for regression problems. It is in this sense that the method is supervised to learn a pre-determined task. Unsupervised methods do not demand a `labelled' dataset and the algorithm tries to find patterns in the data, without labels to supervise or guide. A few applications of supervised learning in astronomy include study of stellar spectra \citep{Bailer-Jones1998,Solorio2005,Sharma2019}, time-domain astronomy \citep{Miller2015,Sedaghat2018,Sanchez2019}, and determination of galaxy morphology \& other properties \citep{Ball2004,Abraham2018}. Unsupervised learning examples include classification of SDSS galaxy spectra \citep{Sanchez2010}, variable stars \citep{Valenzuela2018}, detection of anomalous objects \citep{Rubin2016,Baron2017}, and identification of structures in supernova remnants \citep{Iwasaki2019}. The third category of ML techniques is referred to as reinforcement learning, which is mostly used in the systems where an agent interacts with its environment. In this case, the learning is driven by the rewards/punishments based on the agent's correct/incorrect actions. These kinds of approaches are heavily exploited in robotics, simulations, and game theory \citep{Mnih2013,Baker2016,Silver2016,Vecerik2017}.

Supervised ML approaches have been explored in the past for the stellar spectral interpolation using neural networks and other probabilistic models. \citet{Ness2015} presented a probabilistic generative model named \textit{The Cannon} which learns the mapping from stellar atmospheric parameters (\teff{}, \logg, and \feh) to continuum-normalized stellar spectra. They use this learned mapping to determine the parameters for spectra with unknown properties. The proposed model generates a probability density function with a mean and variance at each wavelength point for each object spectrum from the reference/training set. Their training and test samples contain high-resolution (R$\sim$22,500) spectra from the APOGEE survey \citep{Majewski2017} in IR domain (15200-16900 \AA) with \teff{} coverage of 3500\,-\,5500\,K. The sample mostly consists of giants with only a few dwarf spectra. A machine-learning approach was proposed by \citet{Dafonte2016} to estimate the stellar parameters from the \textit{Gaia} Radial Velocity Spectrograph (RVS) instrument. They used a generative artificial neural network (GANN) for generating stellar spectra from four input stellar parameters, \teff{}, \logg, \feh, and \alfe. For the training and validation of the model, synthetic spectra at medium resolution (R$\sim$11200) were used in the CaII triplet region (847-871 nm) with \teff{} between 4000 to 11500\,K and \logg{} between 2.0 to 5.0 dex.
Another approach that uses a radial basis function (RBF) network for interpolating stellar spectra is introduced in \citet{Cheng2018} where they use the MILES library in the optical domain for training the network using a leave-one-out method. In this approach, the spectrum is interpolated as a linear combination of Gaussian RBFs and the coefficients of the three-dimensional Gaussian function are calculated by solving a set of linear equations. 
Recently \citet{Ting2019} developed a neural network model, \textit{The Payne}, for generating stellar spectra in the APOGEE wavelength domain using 25 stellar ``labels'' which included stellar physical parameters and individual elemental abundances. They train a two layer deep neural network on 2000 synthetic spectra of giants and dwarfs with \teff{} coverage of 3000\,-\,8000\,K. On a test set of 850 synthetic spectra, they obtained rms error of < 0.1\%.

In this study, we consider spectral interpolation as a supervised regression problem and employ Fully-Connected Neural Network (FCNN/ANN/NN) and Random Forest (RF) algorithms specifically designed for handling regression problems. These algorithms have been implemented using Python. We examine ANN and RF algorithms to interpolate the stellar spectra in the optical domain (3500\,-\,7400\,\AA) for low-resolution spectra (R$\sim$2000) with a wider coverage on the temperature axis (\teff{} $\in$ [2800, 35000] K). The training as well as the test set consist of spectra taken from the empirical spectral libraries.

The paper is structured as follows: In Section~\ref{sec:data}, we describe the data used to implement ML algorithms for spectral interpolation. In Section~\ref{sec:method}, we investigate two supervised ML approaches, namely, FCNN and RF, for spectral interpolation and validate our results. In Section~\ref{sec:results}, we discuss the application of the trained model to a sample of spectra from the CFLIB and the LAMOST DR4 v2. Subsequently, we compare our results with those obtained using other existing methods in the literature and examine the outliers. In Section~\ref{sec:conclusion_discussion}, we summarize our results and discuss some further implications of this work.

\section{Data}\label{sec:data}

We aim to develop an ML regression model which can take three stellar atmospheric parameters (\teff{}, \logg{} and \feh{}) as input and provides the corresponding spectrum in the optical wavelength region as an output. For developing such a model using a supervised learning approach, we require a training dataset set which should contain optical stellar spectra with good flux-calibration and corresponding reliable estimates of atmospheric parameters. Other crucial properties that we consider while looking for such datasets are wide coverage of the atmospheric parameters for modelling over the whole parameter space, full optical wavelength coverage without gaps, and sufficiently high signal-to-noise ratio (SNR) spectra with moderate resolution. We find that version 9.1 of the Medium-resolution Isaac Newton Telescope Library of Empirical Spectra \citep[MILES v9.1,][]{miles2006} satisfies all these conditions and can serve as a good training set. The MILES library contains flux-calibrated and normalized spectra (at 5550 \AA) of 985 stars in the wavelength range of 3500\,-\,7429.4 \AA{} with FWHM resolution of 2.56 \AA.

There are two sets of atmospheric parameters available for MILES spectra. The first set comes from \citet{Cenarro2007}, where the estimated parameters using different methods are compiled from various literature sources and brought to a standard reference scale by applying calibrations. Another set of homogeneous determinations is provided by \citet{Prugniel2011} using a polynomial based interpolator, referred to as TGM. They estimate the atmospheric parameters using a full-spectrum fitting technique and establish reliability by comparing their estimates with values from the literature. An updated interpolator, TGM2, was developed in \citet{Sharma2016} to provide improved parameters for 321 cool stars from MILES using the same approach as used in \citet{Prugniel2011}. Availability of two sets of parameters from \citet{Cenarro2007} and \citet{Prugniel2011} plus \citet{Sharma2016} results in two training sets, which we refer to as `Training Set 1' and `Training Set 2' respectively.

For the first training set from \citet[][hereafter CEN]{Cenarro2007}, we find that there are only 946 MILES spectra for which all three parameters are available; therefore the training set prepared based on CEN parameters contains 946 samples. For Training Set 2, we combine the measurements from \citet{Prugniel2011} and \citet{Sharma2016} by adopting the parameters for 321 stars from \citet{Sharma2016} and remaining from \citet{Prugniel2011}. With this, we obtain all three parameters for 984 stars. For star HD 199478 (MILES \# 780), no \logg{} and \feh{} determination is available in either of the two sources and therefore, we exclude this object from the dataset. We refer to this second set of parameters as `PS'. 

Each MILES spectrum has a wavelength range of $\lambda\lambda$ 3500\,-\,7429.4 \AA{} at uniform steps of 0.9 \AA, which results in 4367 flux values. However, we see some spikes at the blue and red ends of MILES spectra. To avoid these spikes while modeling the spectrum, we only consider the region $\lambda\lambda$ 3536\,-\,7410.5\,\AA{}, which gives 4306 flux values. The atmospheric parameters serve as the input features to the model, while flux values are provided as the expected output for the supervised training.

For an independent evaluation of the performance of the trained regression model, we use spectra from CFLIB as the test set. This library contains the spectra of 1273 stars with resolution of 0.88 \AA{} (FWHM) and covers the wavelength range of $\lambda\lambda$ 3465\,-\,9469\,\AA{}. The spectral flux values in the library are provided at 0.4\AA{} wavelength interval. We notice that there are some spectra in the library that contain gaps of more than 50\,\AA{} where flux values are missing. After excluding such spectra, we are left with a sample of 850 stars. To maintain the dimensional and resolution consistency between the training and test set, we pre-process all selected CFLIB spectra through the following procedure:
\begin{enumerate}
\item Degrade the original resolution from 0.88\,\AA{} to 2.56\,\AA{} by convolving with a Gaussian of FWHM\,=\,$\sqrt{2.56^2-0.88^2}$,
\item Resample at every 0.9\,\AA{} by cubic spline interpolation between 3465\,-\,9469\,\AA{},
\item Normalize flux values to 1 at the wavelength point closest to 5550\,\AA,
\item Retrieve flux values corresponding to wavelength range 3536.0\,-\,7410.5\,\AA.
\end{enumerate}
The above steps result in a test set with 4306-dimensional flux values in the same wavelength region and with the same resolution as in the training set. 

For the atmospheric parameters for CFLIB spectra, we again find two sources of measurements from \citet{Valdes2004} and \citet{Wu2011}, which result in two test sets, labelled as `Test Set 1' and `Test Set 2' respectively . Out of 850 selected spectra, we could find the atmospheric parameters for only 649 spectra in \citet{Valdes2004}, whereas \citet{Wu2011} have provided the parameters for all 850 spectra.

\begin{figure}
  \centering
  \includegraphics[width=\linewidth]{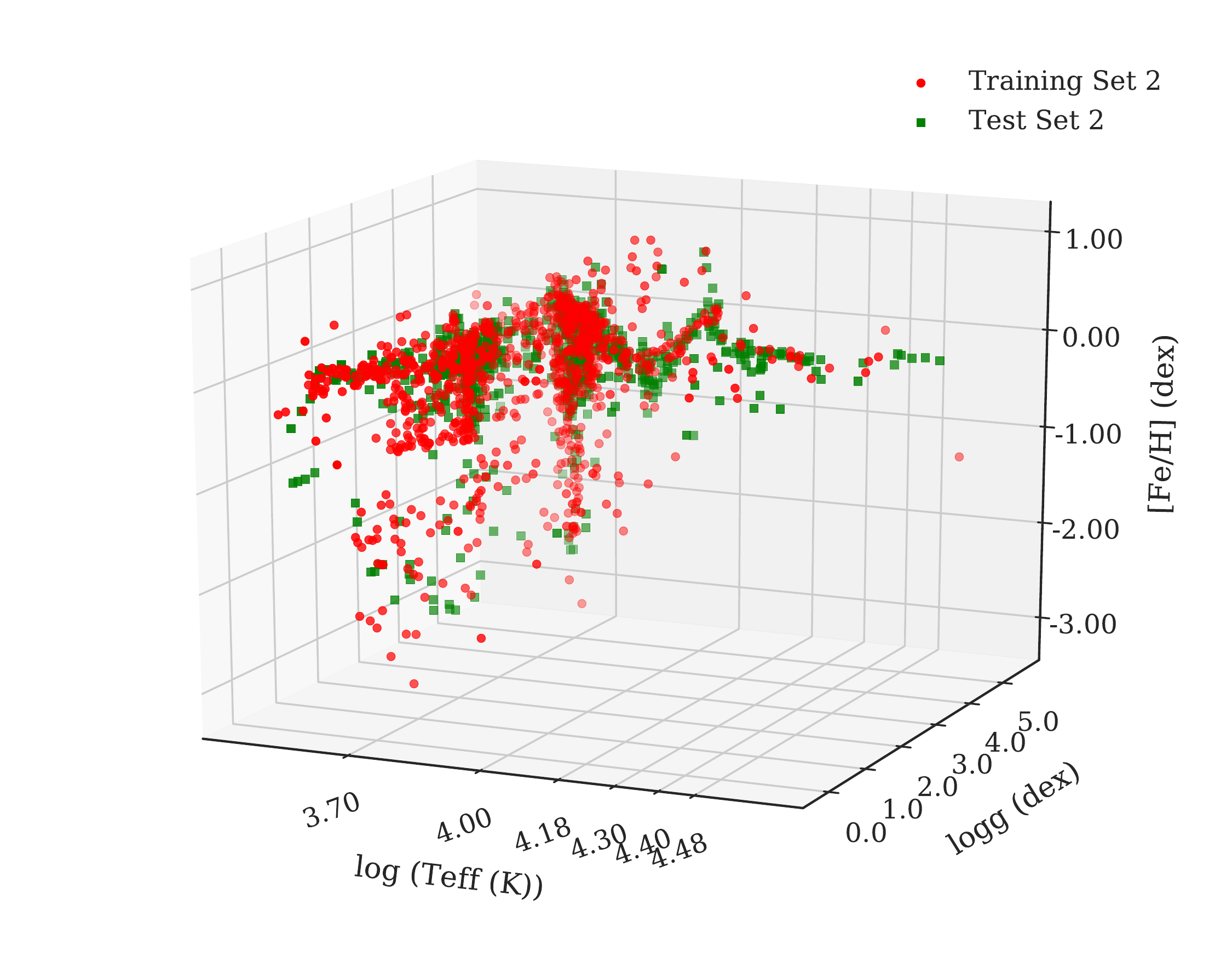}
  \includegraphics[width=\linewidth]{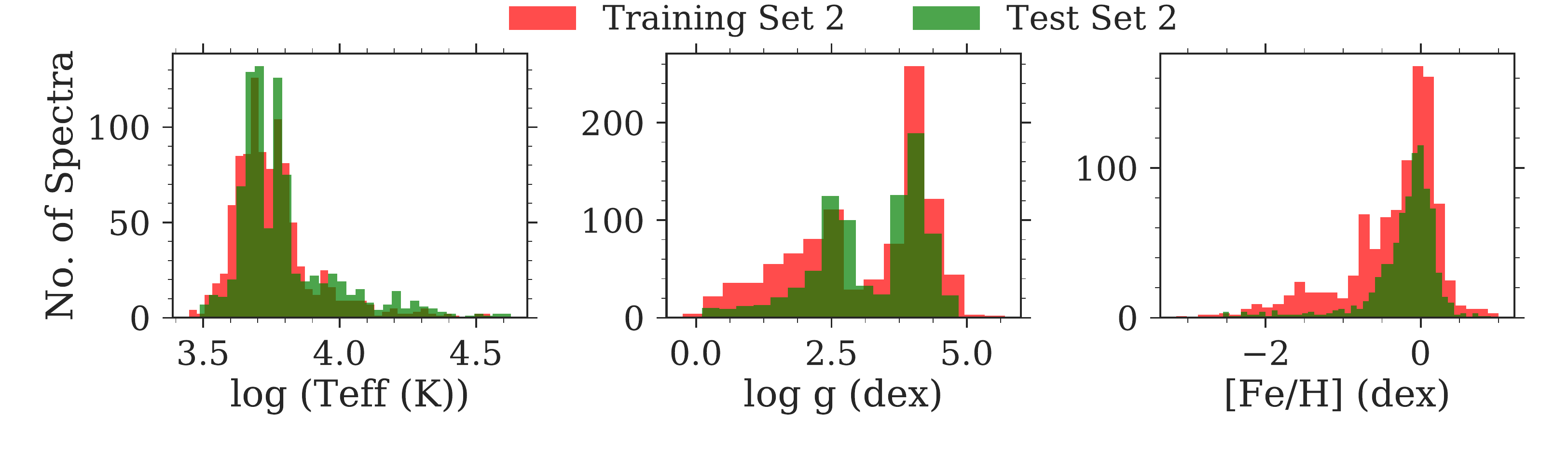}
  \caption{\textit{Top}: Distribution of training and test sets in three-dimensional (\teff, \logg, and \feh) feature space. Training set values are shown as red filled circles, whereas the test set is marked as green squares. \textit{Bottom}: This panel shows the distribution of number of spectra from the training and test sets in different intervals for each of the three parameters separately.}
  \label{fig:params_dist}
\end{figure}

Details of the two training sets and two test sets with the sample size and parameter coverage are provided in Table~\ref{tab:data}. It is noticeable that all datasets span similar parameter space, which can be considered as the range of the parameters over which the final model is applicable. Distribution of the training set 2 and test set 2 (as they have the maximum number of samples) in the parameter space is presented in Fig.~\ref{fig:params_dist}, which shows similar coverage for the training and test sets. The same coverage of feature space for the training and test sets is one of the major criteria for a successful application of machine learning techniques.

\begin{table*}
\caption{Details of selected spectra from MILES for training and from CFLIB for testing the model. For each dataset, the second column contains the number of available spectra with all three parameters and the total number of spectra in the library. The third, fourth, and fifth columns show the coverage over the three-dimensional parameter space. In the last column are the reference papers for the adopted atmospheric parameters.}\label{tab:data}
\centering
\begin{tabular}{lccrcl}
\hline\hline
Database                &     Selected/     & \teff{} Range     & \logg{} Range           & \feh{} Range           & Reference                                \\
                        &    Total stars    &                   &                         &                        &                                          \\
\hline                                    
MILES; Training Set 1   &     946/985       & 2950\,-\,36,000\,K &  $-0.20$\,-\,5.50\,dex  & $-2.86$\,-\,1.65\,dex  & \citet{Cenarro2007}                      \\
MILES; Training Set 2   &     984/985       & 2805\,-\,35,500\,K &  $-0.25$\,-\,5.70\,dex  & $-3.15$\,-\,1.00\,dex  & \citet{Prugniel2011}, \citet{Sharma2016}  \\\hline
CFLIB; Test Set 1       &     649/1273      & 3210\,-\,29,890\,K &   0.00\,-\,5.27\,dex    & $-3.01$\,-\,1.60\,dex  & \citet{Valdes2004}                       \\
CFLIB; Test Set 2       &     850/1273      & 3070\,-\,42,421\,K &   0.04\,-\,5.38\,dex    & $-2.55$\,-\,0.98\,dex  & \citet{Wu2011}                           \\\hline
\hline
\end{tabular}
\end{table*}

\section{Methods}\label{sec:method}

We use two algorithms, Artificial Neural Net (ANN) and Random Forest (RF), for developing the spectral interpolation model. Artificial Neural Networks are inspired by the functioning of the biological neurons, whereas Random Forest is a tree-based regression method, where the final outcome is collectively predicted by different base models, where each base model learns the prediction rules from randomly selected input features.

\subsection{Artificial Neural Net}\label{sec:ann_keras}

We build a fully connected neural network (FCNN) architecture in \texttt{Keras}\footnote{\url{https://keras.io/}} with one input layer which takes the three atmospheric parameters as an input for $n$ sample spectra, two intermediate hidden layers, followed by a final layer which predicts the 4306 flux values as the output. The input parameters are the sequence of [\teff, \logg, \feh] and we note that the values of effective temperature are of the order of $10^3-10^4$ K, whereas other parameters (\logg{} and \feh) are of the order 1. Therefore, we adopt the general approach of standardizing the input features by removing the mean and scaling the residual to unit variance and then feeding these values with zero mean and unit variance to the neural network. Standardizing the input features (atmospheric parameters in our case) brings all the variables within the same numerical range. To deal with the same issue for the output flux values, we normalize each spectrum to unity at 5550 \AA{} by dividing all the flux values by the flux value at $\lambda\,=\,5550$ \AA. This does not make much difference to the MILES library as it is already normalized, but when we consider the spectral interpolation over other libraries which are flux-calibrated in absolute units, this step plays a crucial role.

The number of neurons in the intermediate hidden layers is one of the most important parameters in a neural network algorithm, but there are no specific rules to decide what that number should be in a good model. Other than the number of neurons, there are various ``hyper-parameters'' which play a crucial role in deciding the performance of the model, e.g. the activation function for each neuron, the scheme to initialize the weights, optimization algorithm, etc. Without any standard rule for choosing these parameters, we opt for scanning the hyper-parameter space with the following hyper-parameter grid points:
\begin{Verbatim}[fontsize=\small]
 - neurons1 = [16, 32, 64]
 - neurons2 = [16, 32, 64]
 - optimizer = [`RMSprop', `Adagrad', `Adadelta', `Adam',
                `Adamax', `Nadam']
 - activation = [`softmax', `softplus', `softsign',
                 `relu', `tanh', `sigmoid', `hard_sigmoid',
                 `linear']
 - initializer = [`uniform', `lecun_uniform', `normal',
                `glorot_normal', `glorot_uniform',
                `he_normal', `he_uniform']
\end{Verbatim}
This results in 3024 combinations of hyper-parameters but running the model with each possible hyper-parameter combination is computationally expensive and can take over a couple of days on a system with good GPU capability. Therefore, rather than scanning the whole space, we perform a random search over the hyper-parameter combinations and select 100 combinations randomly. We test the model for these 100 combinations and cross-validate each model five times. The cross-validation is performed to check the generalizability of the model, which is the performance of the model on unseen data not supplied during the training. In the 5-fold cross-validation scheme used here, the whole dataset is split into five groups, only four of which are used for the training. The held-out dataset is used for the validation. This process is repeated for each group as a validation set. After a single run of 5-fold cross validation, the model returns five performance scores, one for each group as validation dataset. The mean and standard deviations in the performance score reflect the performance and generalizability of the model respectively. A model with a high performance score and low standard deviation is considered the best model.

For evaluating the performance of the models in the random grid search and cross-validation, we use average mean squared error (MSE) as the performance metric. We compute the MSE for the $i^{\textrm{th}}$ spectrum as
\begin{eqnarray}\label{eq:mse}
\textrm{MSE}_i & = & \frac{1}{4306}\sum_{j=1}^{4306}(F_{ij} - \hat{F_{ij}})^2,\\
\textrm{Avg. MSE} & = & \frac{1}{n}\sum_{i=1}^{n}\textrm{MSE}_i,
\end{eqnarray}
where $F_{ij}$ and $\hat{F_{ij}}$ are the predicted and expected $j^{\textrm{th}}$ flux values respectively for the $i^{\textrm{th}}$ spectrum. The values of $\textrm{MSE}_i$ in the test set are averaged over $n$ spectra to get average mean squared error, which is used to track the performance of each model over the random combinations of hyper-parameters.

\begin{figure*}
  \centering
  \includegraphics[width=0.8\linewidth]{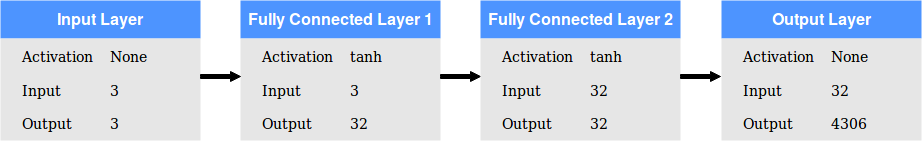}
  \caption{Architecture of the neural network. Three atmospheric parameters (\teff{}, \logg{}, and \feh{}) are supplied to the input layer. The fully connected layers 1 and 2, each have 32 nodes and use the tanh activation function. The final layer takes the input from layer 2 and predicts 4306 flux values.}
  \label{fig:ann_architecture}
\end{figure*}

We find that the combination of 32 neurons in each of the hidden layers with `tanh' activation, `adamax' optimizer and `unform' initializer gives the lowest mean squared error while training on PS parameters. We use these hyper-parameters to set up our final neural net model. The schematic diagram showing the model architecture is presented in Fig.~\ref{fig:ann_architecture}. 

During training the final model presented in Fig.~\ref{fig:ann_architecture}, we set aside 20\% as the data for validation. Since there is a large number of free parameters involved in the training and the training set is not very large, the algorithm is prone to over-fit the data. To avoid this, while training the final model, we use early stopping criteria on the mean absolute error (MAE) for the validation set with the patience level of 50 and minimum delta of 0.00001. This implies that the MAE for the validation set will be monitored for each training epoch and training will stop if there is no improvement in the mean absolute error beyond the set threshold value for the validation sample for 50 consecutive epochs. 

We use MSE (Eq.~\ref{eq:mse}), MAE, and $R^2$ score \citep{Steel1960,Glantz1990,Draper1998} to monitor the training of the model over different epochs. MAE is the average of the absolute values of differences between the original and interpolated spectra. $R^2$ score indicates the proportion of variance in the expected fluxes governed by the predicted fluxes. It can take the values in the range 0-1, and is defined as: 
\begin{eqnarray}\label{eq:r2_score}
R^2 = 1 - \frac{SS_{\textrm{res}}}{SS_{\textrm{tot}}},
\end{eqnarray}
where $SS_{\textrm{res}}$ (sum of square of residuals) and $SS_{\textrm{tot}}$ (total sum of squares) are defined as:
\begin{eqnarray}\nonumber
SS_{\textrm{res}}=\sum _{i}(y_{i}-f_{i})^{2}=\sum _{i}e_{i}^{2};\\\nonumber
SS_{\textrm{tot}}=\sum _{i}(y_{i}-{\bar {y}})^{2}.
\end{eqnarray}
Here $y_i$ and $f_i$ represent the expected and predicted labels and $\bar{y}$ denotes the mean of the expected labels. For a perfect classification model, the value of R$^2$-Score should be equal to 1.

\begin{figure*}
  \centering
  \includegraphics[scale=0.4]{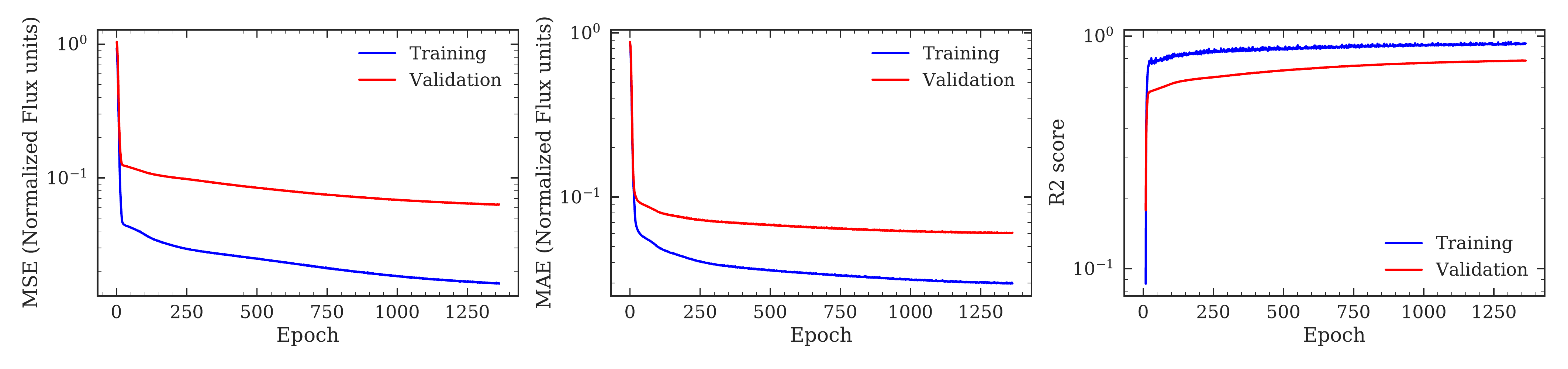}
  \caption{Learning curves for the neural network training on PS parameters. The left, middle, and right panels show the variation in mean squared error, mean absolute error, and $R^2$ score respectively. Although the upper threshold for the number of epochs is 2000, the training stops at an earlier epoch when the validation accuracy saturates and does not improve with further training.}
  \label{fig:val_curve}
\end{figure*}

After each training epoch, the network evaluates the MSE, MAE, and $R^2$ score between the expected and interpolated spectra for the training and validation set. The locus of variations of these quantities over training epochs is called training curve. We show one such training curve for the model trained on PS parameters in Fig.~\ref{fig:val_curve}. The y-axis is in logarithmic scale for better visualization. It is clear that the performance of the model is comparable for training and validation sets. Soon after the first 100 epochs, the mean absolute error drops to less than 0.08 in the normalized flux units for the validation set and decreases further to 0.06 at the final epoch where the training stops. 

\begin{figure*}
  \centering
  \includegraphics[width=\linewidth]{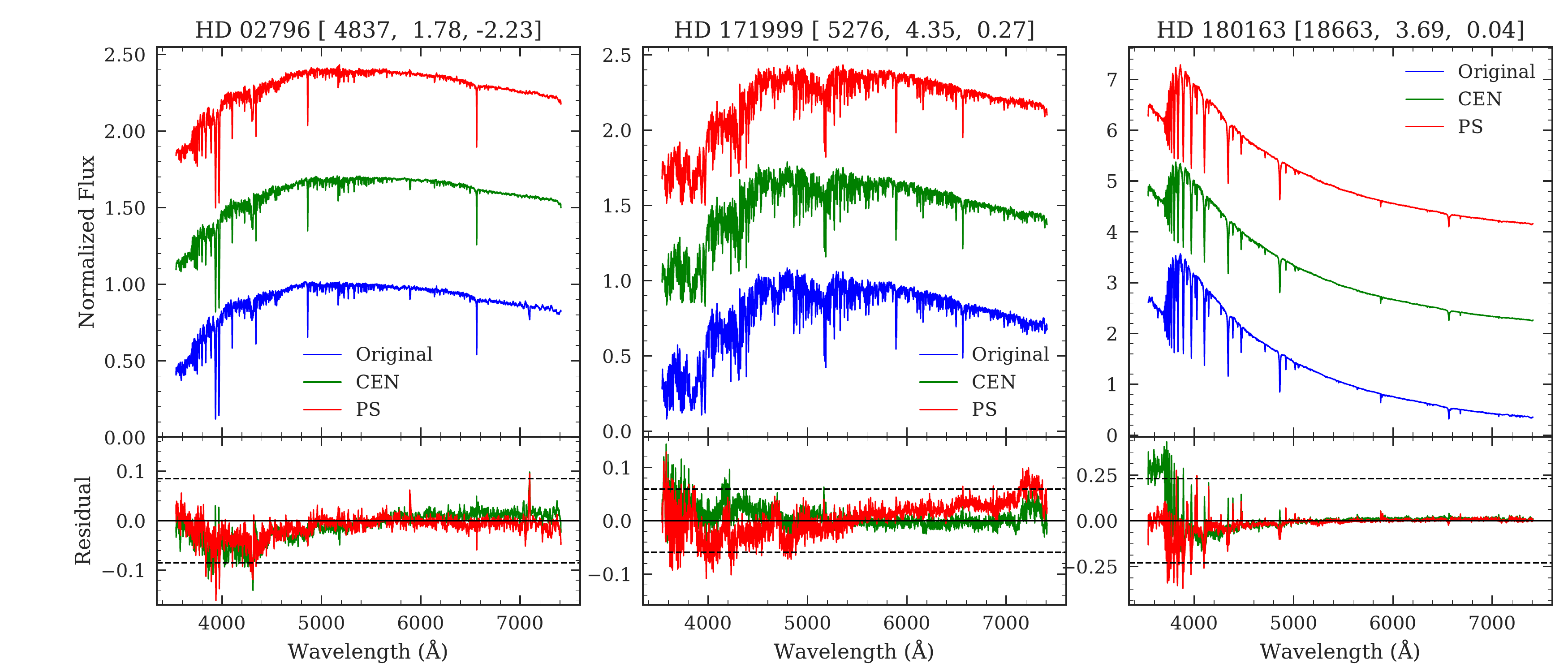}
  \caption{Spectral interpolation using NN for three sample spectra at the edges of the parameter space from the MILES. The upper panels show the original spectra in the blue. Interpolated spectra using the model trained on CEN parameters and PS parameters are shown in green and red color respectively. A vertical offset has been applied to the interpolated spectra for more clarity. Lower panels show the residual (Interpolated-Original) as solid lines and the dashed black lines indicate the $\pm3\sigma$ levels, where $\sigma$ is the standard deviation (Eq.~\ref{eq:sigma}) computed for the residual spectrum obtained from the PS parameters trained model. The spectrum identifier and corresponding atmospheric parameters from the training set 2 are mentioned at the top of each plot.}
  \label{fig:ann_train}
\end{figure*}

We train two separate FCNN models following the procedure described above using training sets 1 and 2 (See Table~\ref{tab:data}). As a first step, we check the performance of the trained models by visually inspecting the original and predicted flux values for three spectra from the training set. The selected spectra are taken from the extreme ends of the parameters space. HD 2796 is a metal-poor star, HD 171999 is a K-type dwarf, and HD 180163 is one of the hottest stars in the sample. Original and reconstructed spectra for these three objects are shown in the upper panels of Fig.~\ref{fig:ann_train} and the residual spectra are shown in the lower panels.
Using residuals for a single spectrum, we compute the standard deviation ($\sigma$) as:
\begin{eqnarray}\label{eq:sigma}
\sigma = \sqrt{\frac{1}{4306}\sum_{j=1}^{4306}(R_{j} - \bar{R})^2},
\end{eqnarray}
where 4306 and $R_j$ are the total number of flux values and $j^{\textrm{th}}$ flux value of the residual spectrum, respectively. $\bar{R}$ denotes the mean of the residuals.

To assess the interpolation using two models quantitatively, we calculate average values of root mean squared errors (RMSE), mean absolute errors (MAE), mean difference, and $R^2$ score for all MILES spectra. These values are reported in Table~\ref{tab:inter_comp}. For a robust estimate of these quantities, we find 3$\sigma$ outliers iteratively and exclude them while computing the statistics. A model trained on CEN parameters and tested on the same set has an average difference equal to $0.0013\pm0.0150$, whereas the other model trained on PS parameters has a mean difference of $0.0015\pm0.0115$. RMSE and MAE values also indicate an overall better performance for the model trained on PS parameters.

\begin{table*}
\caption{Statistics for the comparison of original and interpolated spectra. The original spectra were taken from MILES, CLFIB, and LAMOST, while the interpolated spectra were generated using ML models (NN and RF), and TGM2 interpolator. RMSE, MAE(Mean Absolute Error), and average difference are in normalized flux units and have been computed after removing 3$\sigma$ outliers iteratively for the robust estimation. $R^2$-Score is a unitless quantity. The last column indicates the number of spectra in the test set. TGM2 interpolator is based on the MILES spectra combined with support spectra from external sources.}
\centering
\begin{tabular}{lllccrcl}
\hline\hline
Method                &    Training Set    & Test Set   &    RMSE   & MAE     &  Avg. Difference    & $R^2$ score & N   \\
\hline                                                                        
\multirow{6}{*}{ANN}  &    MILES\_CEN      & MILES      &   0.0320  & 0.0224  & $ 0.0013\pm0.0150$  & 0.9835   & 946 \\
                      &    MILES\_PS       & MILES      &   0.0285  & 0.0197  & $ 0.0015\pm0.0115$  & 0.9876   & 984 \\
                      &    MILES\_CEN      & CFLIB      &   0.0618  & 0.0395  & $-0.0123\pm0.0248$  & 0.9557   & 850 \\
                      &    MILES\_PS       & CFLIB      &   0.0628  & 0.0395  & $-0.0126\pm0.0268$  & 0.9605   & 850 \\
                      &    MILES\_CEN      & LAMOST     &   0.0495  & 0.0361  & $ 0.0049\pm0.0113$  & 0.9351   & 7   \\
                      &    MILES\_PS       & LAMOST     &   0.0495  & 0.0368  & $ 0.0045\pm0.0107$  & 0.9370   & 7   \\\hline
\multirow{6}{*}{RF}   &    MILES\_CEN      & MILES      &   0.0137  & 0.0097  & $-0.0002\pm0.0074$  & 0.9967   & 946 \\
                      &    MILES\_PS       & MILES      &   0.0127  & 0.0093  & $ 0.0001\pm0.0067$  & 0.9973   & 984 \\
                      &    MILES\_CEN      & CFLIB      &   0.0652  & 0.0417  & $-0.0145\pm0.0264$  & 0.9524   & 850 \\
                      &    MILES\_PS       & CFLIB      &   0.0646  & 0.0392  & $-0.0123\pm0.0261$  & 0.9568   & 850 \\
                      &    MILES\_CEN      & LAMOST     &   0.0450  & 0.0337  & $ 0.0031\pm0.0082$  & 0.9459   & 7   \\
                      &    MILES\_PS       & LAMOST     &   0.0509  & 0.0385  & $ 0.0053\pm0.0124$  & 0.9330   & 7   \\\hline
TGM2                  &    MILES           & MILES      &   0.0274  & 0.0196  & $ 0.0010\pm0.0107$  & 0.9905   & 1020\\
TGM2                  &    MILES           & CFLIB      &   0.0873  & 0.0538  & $-0.0136\pm0.0672$  & 0.8490   & 850 \\
TGM2                  &    MILES           & LAMOST     &   0.0514  & 0.0382  & $ 0.0085\pm0.0109$  & 0.9309   & 7   \\
\hline\hline                                                                        
\end{tabular}\label{tab:inter_comp}
\end{table*}

\subsection{Random Forest}\label{sec:rf}

Random forest \citep[RF;][]{Breiman2001} is an ensemble method for classification and regression problems where an ensemble of base models, referred to as `trees', make independent predictions and the final prediction is made by aggregating the output from each base model. During the training of the RF model, a set of decision rules is extracted by creating multiple tree structures. Each tree structure starts with a root node which is split into subsequent branches using the information contained in the trained dataset and goes down to the final leaf nodes where the predictions are made. Based on the predictions from the leaf nodes, a final value is assigned to the input features. Features considered in each tree for finding the decision rules is a random subset of all features in the dataset and the final outcome is the average from all such randomly selected trees developed using random features selection. Therefore, the problem of over-fitting is inherently checked to some extent within the model.

We use the RF regression to interpolate stellar spectra labelled by \teff, \logg, and \feh{}. The training and test datasets used for RF regression are the same as those used for neural nets and detailed in Table~\ref{tab:data}. For the training, we use ten estimators (number of trees) with mean squared error as the criterion to measure the quality of the split. Apart from the number of estimators and split-criterion, we also optimize the other hyper-parameters of the algorithm and find that the different hyper-parameter settings do not result in substantially different results. Therefore, we decide to use the default values for the other hyper-parameters. For implementing the RF regression, we use the python machine-learning library \texttt{scikit-learn}\footnote{\url{https://scikit-learn.org/stable/index.html}}\citep{scikit-learn}.

\begin{figure*}
  \centering
  \includegraphics[width=\linewidth]{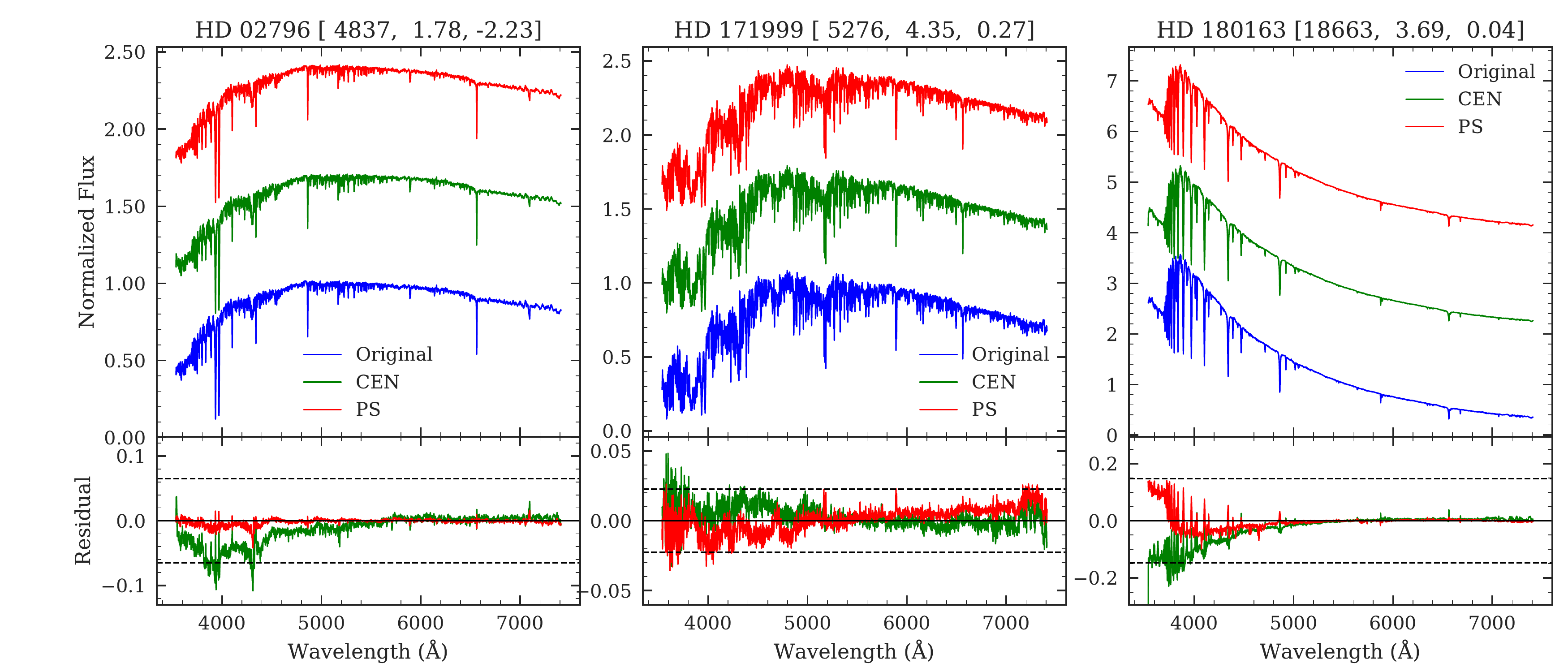}
  \caption{Spectral interpolation using RF for three sample spectra at the edges of the parameter space from the MILES. Upper panels show the original spectrum in the blue. Interpolated spectra using the model trained on CEN parameters and PS parameters are shown in green and red color respectively. A vertical offset has been applied to the interpolated spectra for more clarity. Lower panels show the residual (Interpolated-Original) as solid lines and the dashed black lines indicate the $\pm3\sigma$ levels, where $\sigma$ is the standard deviation (Eq.~\ref{eq:sigma}) computed for the residual spectrum obtained from the PS parameters trained model. Spectrum identifier and the corresponding atmospheric parameters from the training set 2 are mentioned at the top of each plot.}
  \label{fig:rf_train}
\end{figure*}

Using the RF technique and the two training sets results in two RF models. For the visual inspection of the quality of the spectral interpolation, we present in Fig.~\ref{fig:rf_train} the three spectra discussed for ANN models. Regression using RF models matches reasonably well with the original spectra. For an overall quantitative assessment of the interpolation, we compute previously used performance metrics (RMSE, MAE, Avg. difference, and $R^2$ Score) indicating the quality of the regression and report them in Table~\ref{tab:inter_comp}. 

Based on the performance metrics used to assess the quality of the models, listed in Table~\ref{tab:inter_comp}, we see that the RF model trained on PS parameters achieves better accuracy in generating stellar spectrum based on the three atmospheric parameters as compared to the model trained on CEN parameters. The statistics also suggest that RF regression models outperform NN models in this exercise.

It is important to note that we have assessed the performance so far using spectra which the network has already seen. For an independent assessment of the performance of each individual ML model, we use a separate spectral library, CFLIB, as the test set. The results are presented in the next section, Sec.~\ref{sec:results}. 

\section{Results}\label{sec:results}

For an independent analysis of the performance of the NN and RF models, we apply these models to the test set compiled from the CFLIB which was not part of the training set. There are two sets of measurements of atmospheric parameters available for the CFLIB spectra. \citet{Valdes2004} compiled the atmospheric parameters (`Test Set 1') for CFLIB stars from various independent studies. In another study by \citet{Wu2011}, the authors have homogeneously determined the parameters of all the spectra from the CFLIB using a single technique (`Test Set 2'). However, noting that there might be systematic differences in the parameters obtained from different methods of determination as in the case with \citet{Cenarro2007}, we adopt the parameters from the latter study to generate the spectra. 

\begin{figure}
  \centering
  \includegraphics[width=0.85\linewidth]{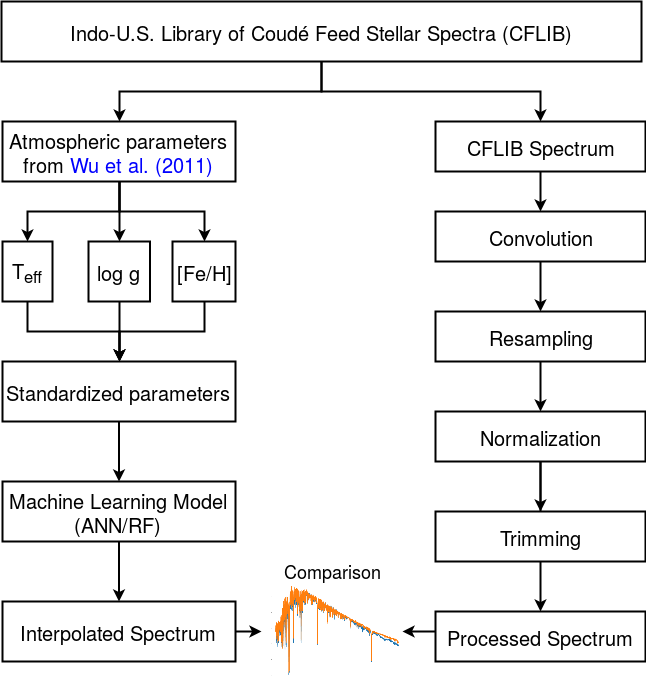}
  \caption{Schematic diagram showing the process of testing the trained ML model.}
  \label{fig:comparison_block_dia}
\end{figure}

Using ML models trained on PS parameters, stellar spectra for the parameters in \citet{Wu2011} are generated. These spectra are compared with the corresponding processed CFLIB spectra, where the pre-processing steps remain the same as described in Sec.~\ref{sec:data}. The schematic diagram showing the process of testing the trained models is presented in Fig.~\ref{fig:comparison_block_dia}.  

\begin{figure*}
\centering
\includegraphics[scale=0.24]{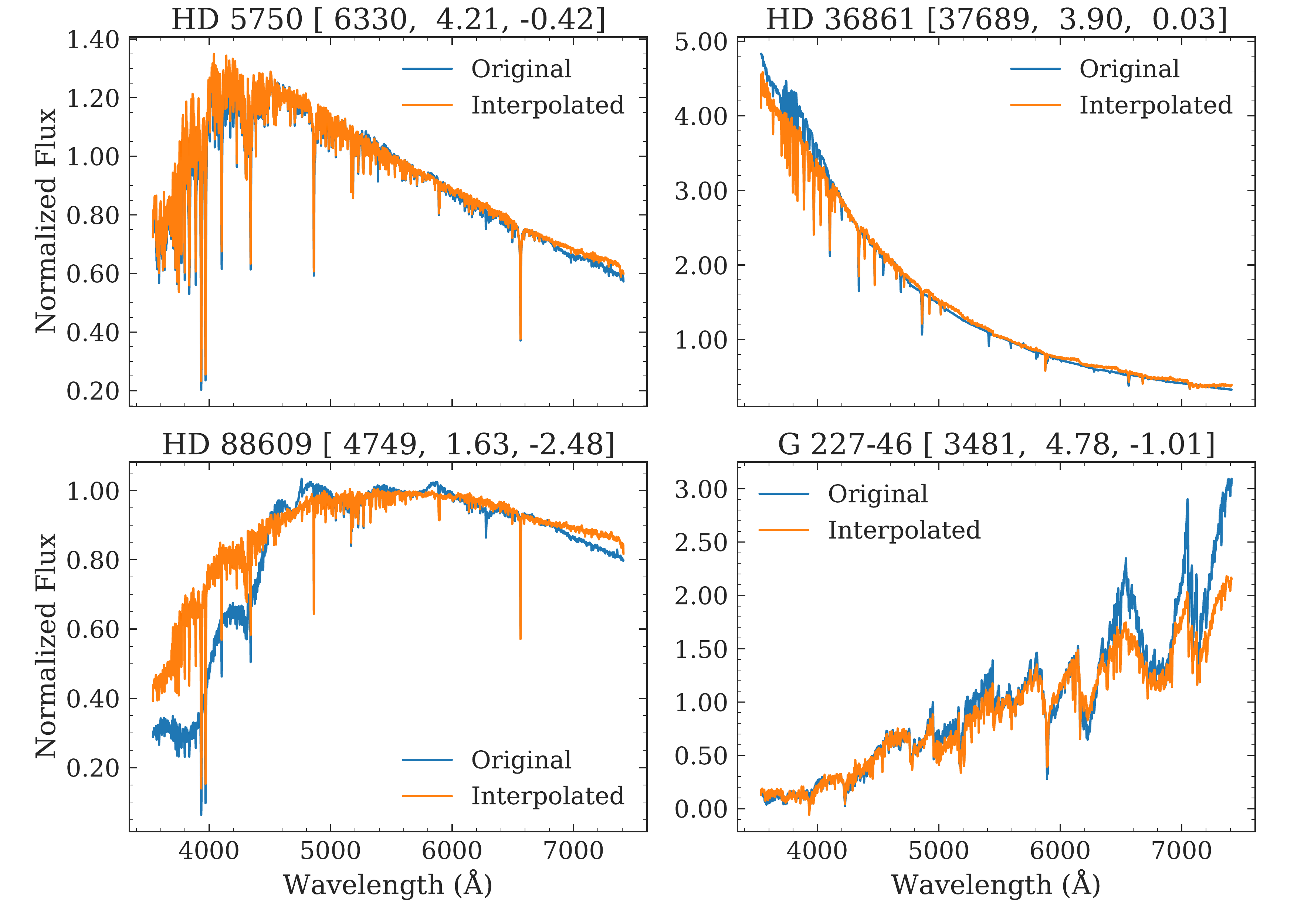}
\includegraphics[scale=0.24]{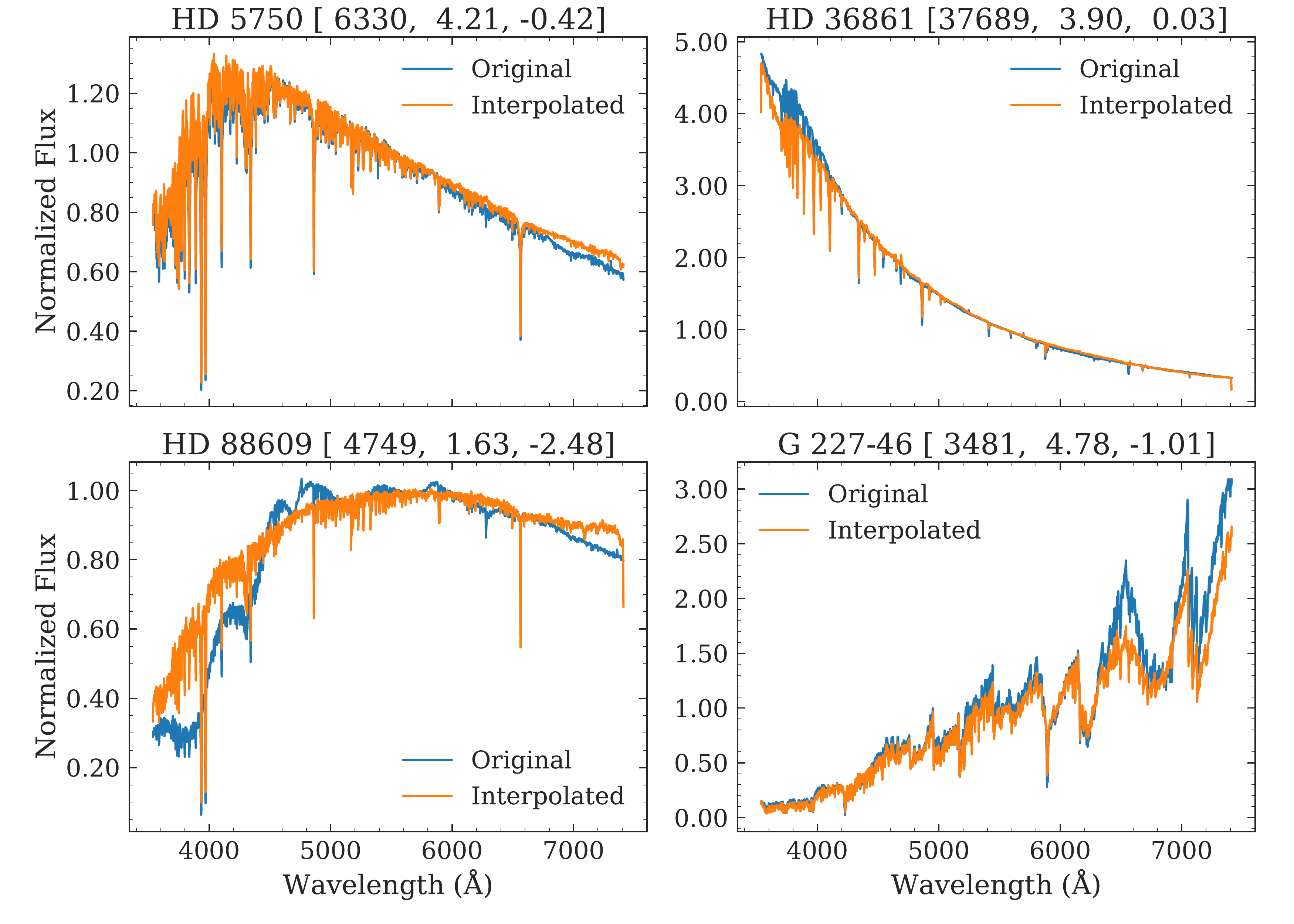}
\caption{Comparison of four original spectra from CFLIB and generated spectra using NN (left panel) and RF model (right panel) trained on MILES. Blue color represents original CFLIB spectra and orange represents spectra generated using the ML model based on atmospheric parameters from \citet{Wu2011}.}
 \label{fig:cflib_sample}
\end{figure*}

For a visual comparison, generated spectra for four sample stars using NN and RF are shown in Fig.~\ref{fig:cflib_sample}. Reconstructed spectra using NN and RF models are shown in the left and right panels respectively for better relative comparison between the two models. 
We notice that the reconstructed spectra for HD 5750 and HD 36861 match reasonably well with the original spectra. However, for HD 88609, the match is not satisfactory. This object is also a part of MILES and comparing the CFLIB spectrum with the MILES spectrum reveals that there is flux-calibration issue in the original CFLIB spectrum. For G 227-46, the interpolated spectrum is underestimated as compared to the original spectrum in the higher wavelength region. For the first three stars, the NN and RF models perform similarly but for the cool star, G 227-46, the spectrum generated using the RF model is closer to the original one as compared to the NN model. 

\begin{figure}
\centering
\includegraphics[scale=0.5]{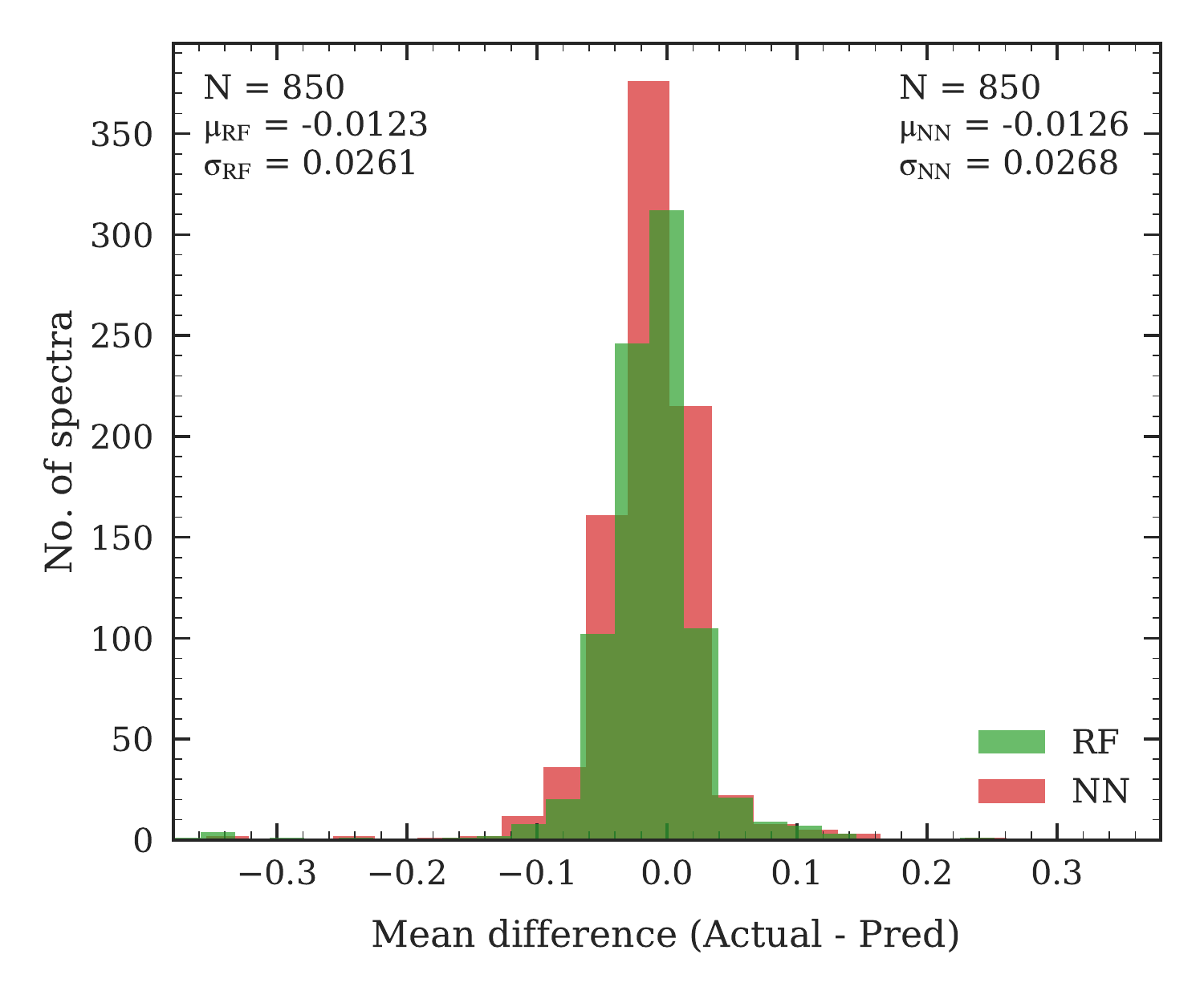}
\caption{Histogram for the mean difference between the actual and predicted flux values using RF (green) and NN model (red) for 850 CFLIB spectra. The interpolated spectra were obtained using the models trained over training set 2. The statistics for both the models are mentioned in the panel where `$\mu$' is the average difference between the original and interpolated spectra and `$\sigma$' denotes the standard deviation.}
 \label{fig:diff_hist}
\end{figure}

For a quantitative assessment of the comparison between original and generated CFLIB spectra, we compute the usual error statistics reported in Table~\ref{tab:inter_comp}. On the basis of the three error indicators, RMSE, MAE, Avg. difference and $R^2$ score, the two ML models are comparable and all the error values are consistent within uncertainties. By comparing all CFLIB original spectra with NN model generated spectra, we get an average RMSE of 0.0628, MAE of 0.0395 with the mean difference of $-0.0126\pm0.0268$. On the other hand, the RF model generated spectra give an average RMSE of 0.0646 and MAE of 0.0392 with 3$\sigma$-clipped mean difference of $-0.0123\pm0.0261$. There is no significant difference in the overall performance of the two models. Histograms of the mean difference between original and interpolated CFLIB spectra using the two models are presented in Fig.~\ref{fig:diff_hist}. From the histograms too, it is clear that the two distribution are similar and the mean difference values are very well centered close to 0. We also model the test set spectra using NN and RF models trained on CEN parameters and report the performance metrics in Table~\ref{tab:inter_comp}.

One key advantage with the Neural Nets is their transfer learning capabilities. Transfer learning is the domain of machine learning/deep learning applications where existing network with pre-adjusted weights are fine-tuned further for another similar application \citep{Gupta2016}. It is easier to fine-tune the weights of a pre-trained NN based spectral interpolation model if it has to be adopted for other type of stars, e.g. chemically peculiar stars, variable stars, etc. Moreover, as shown in Sec.~\ref{sec:ann_keras}, there are standard procedures to track the training of a NN model and stop it when it starts over-fitting. This is important as over-training of a ML based spectral interpolator may lead to its strange behaviour in the less populated regions which might result in an inconsistent spectrum in those regions. This aspect becomes more relevant if these models are to be used for the extrapolation (predicting spectrum beyond the parameter space volume occupied by the training set). In our case, we find that the two models, NN and RF, are equally good based on their global performance metrics (average values of RMSE, MAE, and mean difference) for the test set. However, we decide to adopt RF model over NN model as our preferred model. The primary reason to consider RF over NN is its better performance at the edges of the parameters space (e.g., cool stars) where the density of the training sample is scarce. Also, the model training time is much lesser for RF than for the NN model. On a CPU machine equipped with 8 GB RAM and Intel Core i7-4790 processor with eight cores, training the RF model takes about five seconds as opposed to 263 seconds taken by NN model training for more than 1300 epochs. Time taken for generating the spectra of about 1000 stars is comparable for both the models (2\,-\,3 seconds). It is worth recalling here that the two models use different approaches for the training and the NN implementation was in Keras, whereas the RF implementation uses \textit{scikit-learn} package.

\begin{figure*}
  \centering
\includegraphics[scale=0.46]{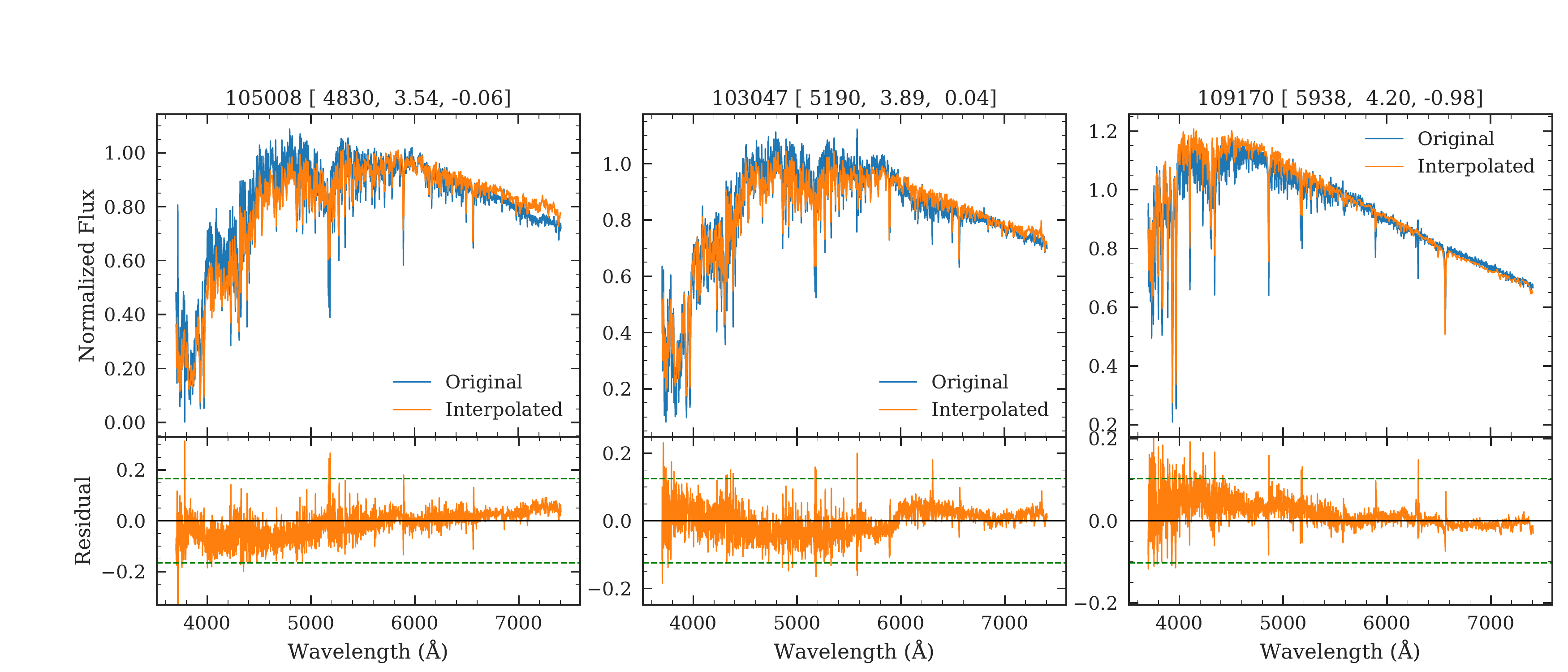}
\caption{Comparison of three LAMOST spectra with interpolated spectra. LAMOST ID and corresponding atmospheric parameters from the LAMOST pipeline are indicated in the title of each plot. \textit{Top panels:}Blue color represents original LAMOST spectra and orange represents the spectra generated using LAMOST parameters with the RF model trained on MILES spectra with PS parameters. \textit{Bottom panels:} Residual (interpolated$-$original) fluxes are shown in orange color with the dashed green lines indicating $\pm3\sigma$ levels, where $\sigma$ denotes the standard deviation (Eq.~\ref{eq:sigma}) computed for a given residual spectrum.}
 \label{fig:lamost_comp_ANN}
\end{figure*}

To examine the capability of the finally adopted RF model, we further apply it on seven LAMOST spectra randomly taken from LAMOST DR4 v2 archive \footnote{\url{http://dr4.lamost.org/}} along with their estimated parameters from the LAMOST Stellar Parameter Pipeline. We generate the spectra for the estimated parameters. Since the spectral resolution of the interpolated spectra is the same as the resolution of the MILES spectra and the LAMOST spectra have lower resolution (FWHM resolution\,=\,3.05 \AA), we convolve the generated spectra with a Gaussian kernel of suitable FWHM before making a comparison. Also, LAMOST spectral coverage starts from $\sim$3700 \AA{} and extends beyond 9000 \AA, but the generated spectra span the wavelength range of 3536\,-\,7410.5 \AA. Therefore, we clip the original and interpolated spectra between the common wavelength region, 3700\,-\,7410.5 \AA. For visual comparison, we show the three LAMOST spectra in Fig.~\ref{fig:lamost_comp_ANN} along with the generated spectra (top panels) and provide a quantitative analysis of the comparison in Table~\ref{tab:inter_comp}.

\subsection{Outliers}\label{sec:outliers}

\begin{figure}
  \centering
\includegraphics[scale=0.48]{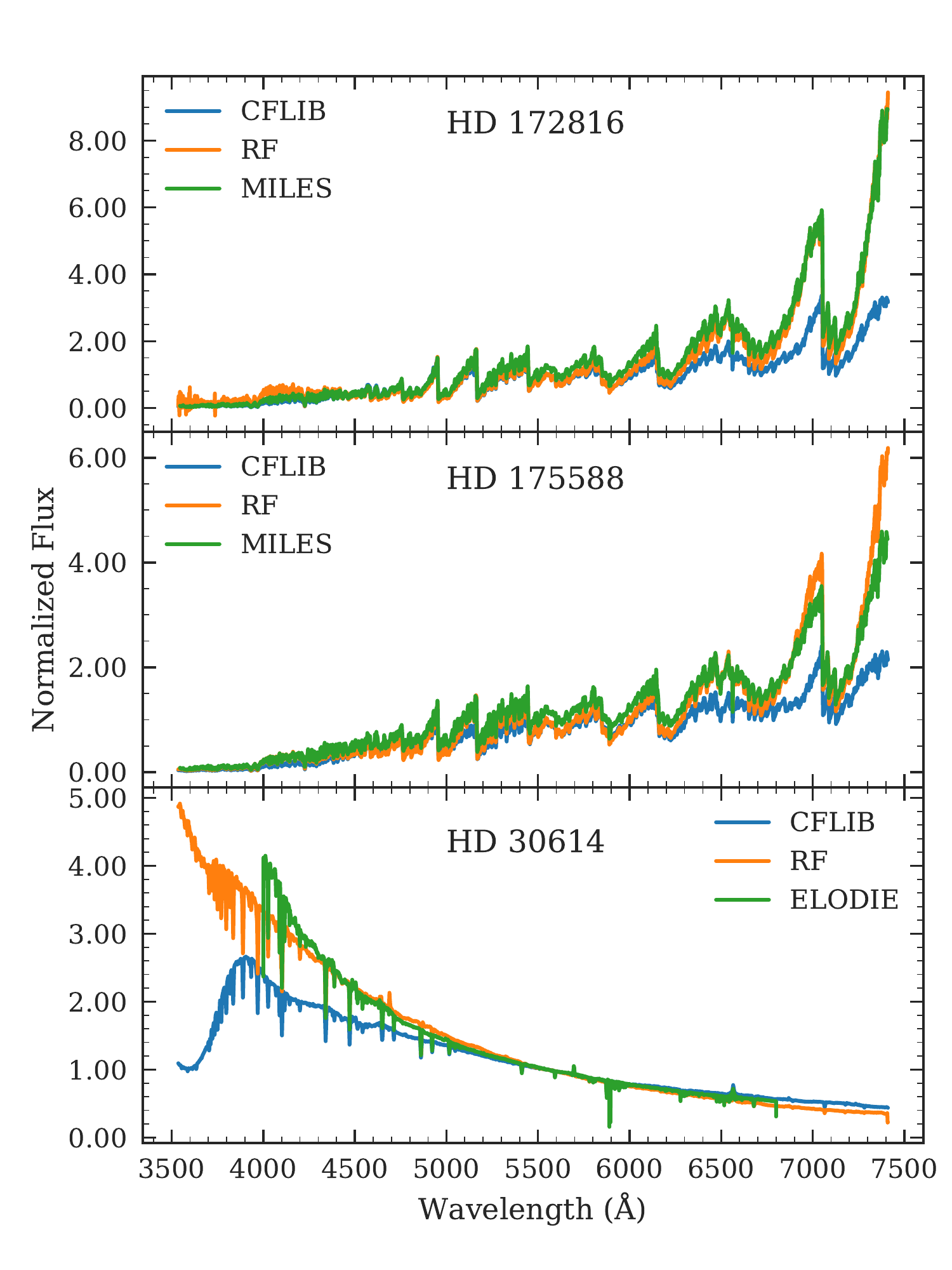}
\caption{Comparison of CFLIB spectra with spectra retrieved from other sources (MILES/ELODIE) for discrepant cases (Table~\ref{tab:outliers}). Each panel also shows the RF model generated spectrum.}
 \label{fig:flux_calib_issue}
\end{figure}

We further study those individual CFLIB spectra where the mean absolute error between the original and reconstructed spectrum is greater than 3$\sigma$. There are 14 such cases ($\sim$\,2\%), listed in Table~\ref{tab:outliers}. Eight of these stars are M giants and the remaining six stars are hotter dwarfs with near-solar to sub-solar metallicity (\feh{}\,<\,0.0). We first check the density of cool (\teff{}\,<\,3500\,K) giants in the training parameter space and find that there exist only six spectra within this region; this results in inadequate training and consequently higher error in generating the spectra from the atmospheric parameters. We also find that some spectra in the CFLIB suffer from the flux-calibration issues \citep{Wu2011} as the spectra were flux-calibrated using the template spectrum of closest matching spectral type from \citet{Pickles1998} library. We show one such CFLIB spectrum for HD 88609 with erroneous flux-calibration in the lower left panel of Fig~\ref{fig:cflib_sample}. This is a G5 type star but the irregular continuum does not match with the spectrum of a typical G5 type star. To dissect the improper flux-calibration aspect, we check the availability of the spectra for the stars listed in Table~\ref{tab:outliers} in other spectral libraries. We find that HD 88609, HD 172816, HD 123657, and HD 175588 are part of MILES too. We present the spectra for HD 172816 and HD 175588 in Fig~\ref{fig:flux_calib_issue}. There is a significant difference in the fluxes between MILES and CFLIB spectra towards the redder wavelengths. We also checked the ELODIE archive \footnote{\url{http://atlas.obs-hp.fr/elodie/}} and found one or more spectra for five common stars: HD 84748, HD 123657, HD 175588, HD 57651, and HD 30614. We could not use ELODIE spectra to verify the flux-calibration issues for cool stars due to their lower signal-to-noise ratio and lack of coverage beyond 6800 \AA{} where the discrepancy dominates. However, the ELODIE spectrum for HD 30614, the hottest star among the outliers, is used to test the spectral energy distribution (SED) of the CFLIB spectrum. This is plotted in the lower panel of Fig.~\ref{fig:flux_calib_issue}. It is evident from the plot that there is a significant difference between the continuum level of CFLIB and MILES spectra and the RF model generated spectra are closer to the actual spectrum. 

\begin{table}
\caption{CFLIB Spectra with mean absolute error greater than 3$\sigma$. The atmospheric parameters are from \citet{Wu2011}.}
\centering
\begin{tabular}{lrrrr}
\hline\hline
 Identifier &  \multicolumn{1}{c}{Teff} &  logg &  Fe/H & \multicolumn{1}{c}{MAE} \\
\hline
   HD 84748 &   3070.0 &  0.78 & -1.00 &  0.8086 \\
   HD 78712 &   3101.0 &  0.23 & -1.00 &  0.3638 \\
  HD 197812 &   3119.0 &  0.32 & -1.00 &  0.5179 \\
  HD 172816 &   3155.0 &  0.14 & -0.44 &  0.4068 \\
  HD 206632 &   3197.0 &  0.33 & -0.30 &  0.3505 \\
  HD 123657 &   3235.0 &  0.48 & -0.20 &  0.3350 \\
  HD 175588 &   3333.0 &  0.48 & -0.04 &  0.2933 \\
   HD 57651 &   3418.0 &  0.85 & -0.07 &  0.2402 \\\hline
  HD 162570 &   7511.0 &  3.87 &  0.02 &  0.4698 \\
  HD 154660 &   7663.0 &  3.97 & -0.18 &  0.4612 \\
 BD+45 1668 &   8693.0 &  4.72 & -0.48 &  0.4245 \\
   HD 29763 &  10073.0 &  2.60 & -0.51 &  0.2311 \\
  HD 179588 &  11165.0 &  4.05 &  0.00 &  0.3546 \\
   HD 30614 &  32591.0 &  3.17 & -0.10 &  0.3557 \\
\hline\hline
\end{tabular}\label{tab:outliers}
\end{table}

\begin{figure}
  \centering
\includegraphics[scale=0.48]{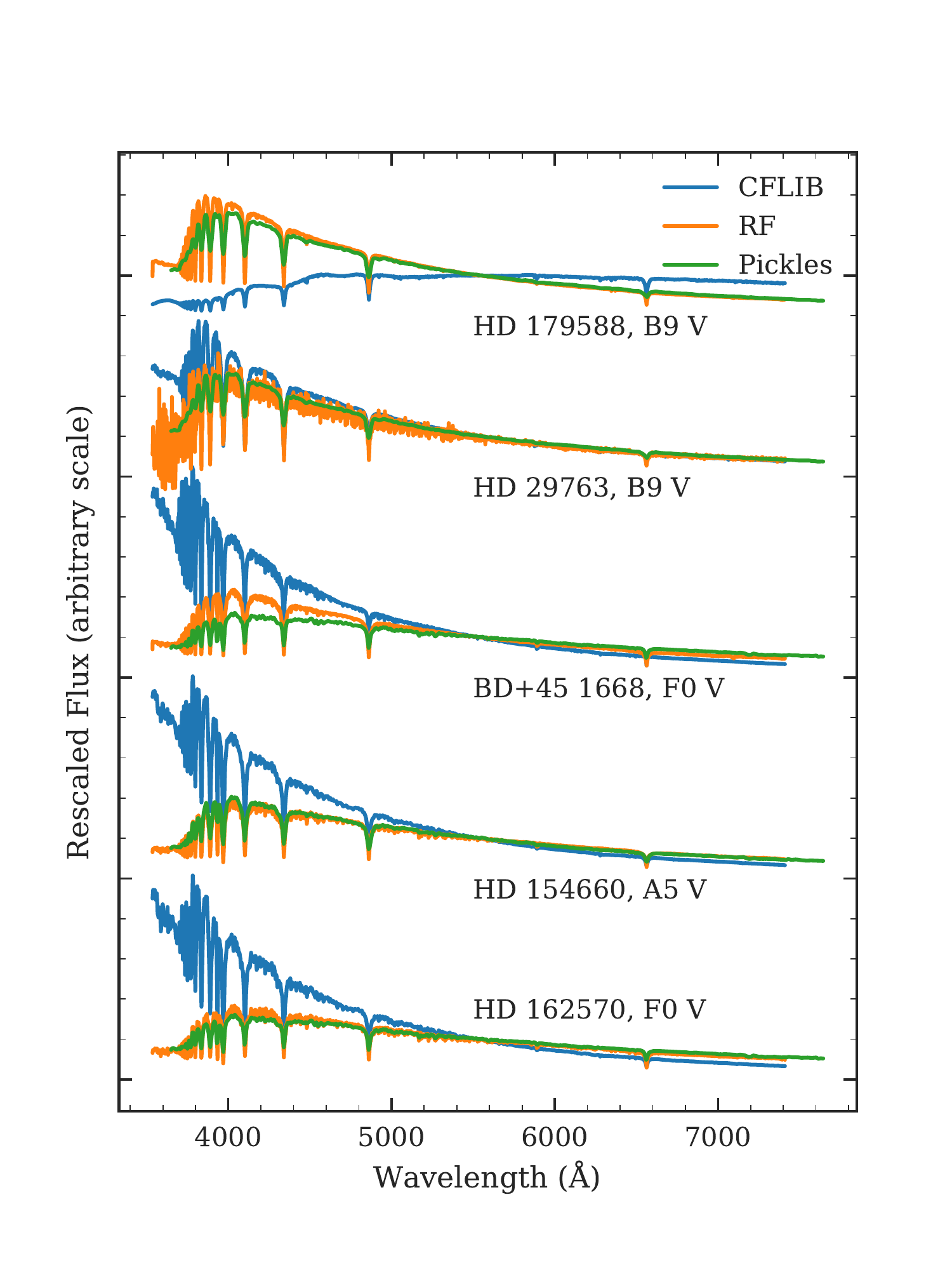}
\caption{Comparison of CFLIB spectra with template spectra from Pickles library for discrepant cases (Table~\ref{tab:outliers}) with \teff{}\,>\,7500\,K. RF model generated spectra are also shown for each case. Spectra corresponding to different sources are vertically shifted by an arbitrary value to avoid overlap. Each spectrum identifier is followed by the closest spectral type of the template found in the Pickles library and used for examining the flux-calibration issue.}
 \label{fig:flux_calib_issue_hotter}
\end{figure}

For discrepant cases hotter than \teff{}\,>\,7500\,K, we check the spectral type from the literature sources and compare the CFLIB spectrum with the closest matching template from \citet{Pickles1998} library and RF model generated spectrum in Fig.~\ref{fig:flux_calib_issue_hotter}. We observe that in all the cases, there is a large difference between the template and CFLIB spectrum. In contrast, the RF model generated spectra show better agreement with the template spectra. Out of six outliers hotter than 7500\,K, four stars (HD 154660, HD 162570, HD 29763, and BD+45 1668) are also discussed in \citet{Sharma2019} as misclassified spectra. They conclude that the most compelling reason for the misclassification is the flux-calibration problem. We, too, believe that the higher error in the interpolated spectrum for these spectra is due to the flux-calibration issues in the CFLIB data and is not associated with any irregularity in the ML models.

\subsection{Comparison with other approaches}\label{sec:comparison}

One of the existing approaches for generating a stellar spectrum from the three atmospheric parameters, \teff, \logg, and \feh, is polynomial based interpolation. TGM2 \citep{Sharma2016} is one such interpolator that approximates the spectrum in each wavelength bin with polynomials in the three parameters. To obtain the coefficients of the polynomials which minimize the difference between the interpolated and observed spectra from a reference spectral library, an input catalogue is prepared. The input catalogue contains the compiled parameters from the literature for the reference library stars. The TGM2 interpolator is primarily based on the MILES spectral library. Some spectra from the other sources were also included to provide extrapolation support at the edges of the parameter space and to populate the sparse regions in the parameter space (for more details see \citet{Prugniel2011}). Combining the reference library spectra with the spectra from external sources results in 1020 spectra which are used for preparing the interpolator. We use TGM2 to interpolate the spectra used in the input catalogue. This is equivalent to testing a trained model on the same sample as was used in the training stage. Comparing the two sets of spectra, we obtain an RMSE value of 0.0274 with MAE of 0.0196. The statistics is presented in Table~\ref{tab:inter_comp}. 

We also generate spectra for 984 MILES stars using TGM2 and PS parameters. We measure the disagreement between the original and interpolated spectra using the same statistical quantities: RMSE, MAE, Avg. difference, and $R^2$ score. Using TGM2, we obtain an average RMSE of 0.0277 in the normalized flux units as compared to the value of 0.0285 using a NN model and 0.0127 using a RF model. Similarly, TGM2 gives MAE of 0.0200, whereas NN and RF give 0.0197 and 0.0093 respectively. $R^2$ scores and average difference values also imply that the performance of the NN model and TGM2 on the training set are comparable but the RF model performs the best on all statistical parameters. 

To assess the true performance of the interpolator, as in the cases of NN and RF, we use TGM2 to interpolate 850 CFLIB spectra. We obtain average RMSE equal to 0.0873 for the difference between the library and interpolated spectra with MAE equal to 0.0538. The error statistics presented in Table~\ref{tab:inter_comp} shows that the TGM2 errors are larger than those obtained using NN and RF models.

\citet{Cheng2018} demonstrate another implementation of an automated approach for interpolating the stellar spectra. They consider Gaussians as the basis functions and use the linear combination of Gaussians as the interpolating function where the parameters of Gaussians are obtained by minimizing the distance between the library spectrum and the interpolated spectrum. We use their RBF interpolation script and interpolate 850 test spectra from the CFLIB. With this model, we obtain the values of RMSE, MAE and Avg. difference equal to 0.0825, 0.0543, and $-0.0158\pm 0.0348$ respectively. The error values are comparable to those obtained from TGM2 interpolator. Though, with RBF interpolator, we get an $R^2$ score of 0.9067 which is better than that from the TGM2 interpolator. However, the error values with RBF interpolator are larger as compared to ANN and RF model.

\begin{figure*}
  \centering
        \includegraphics[width=\linewidth]{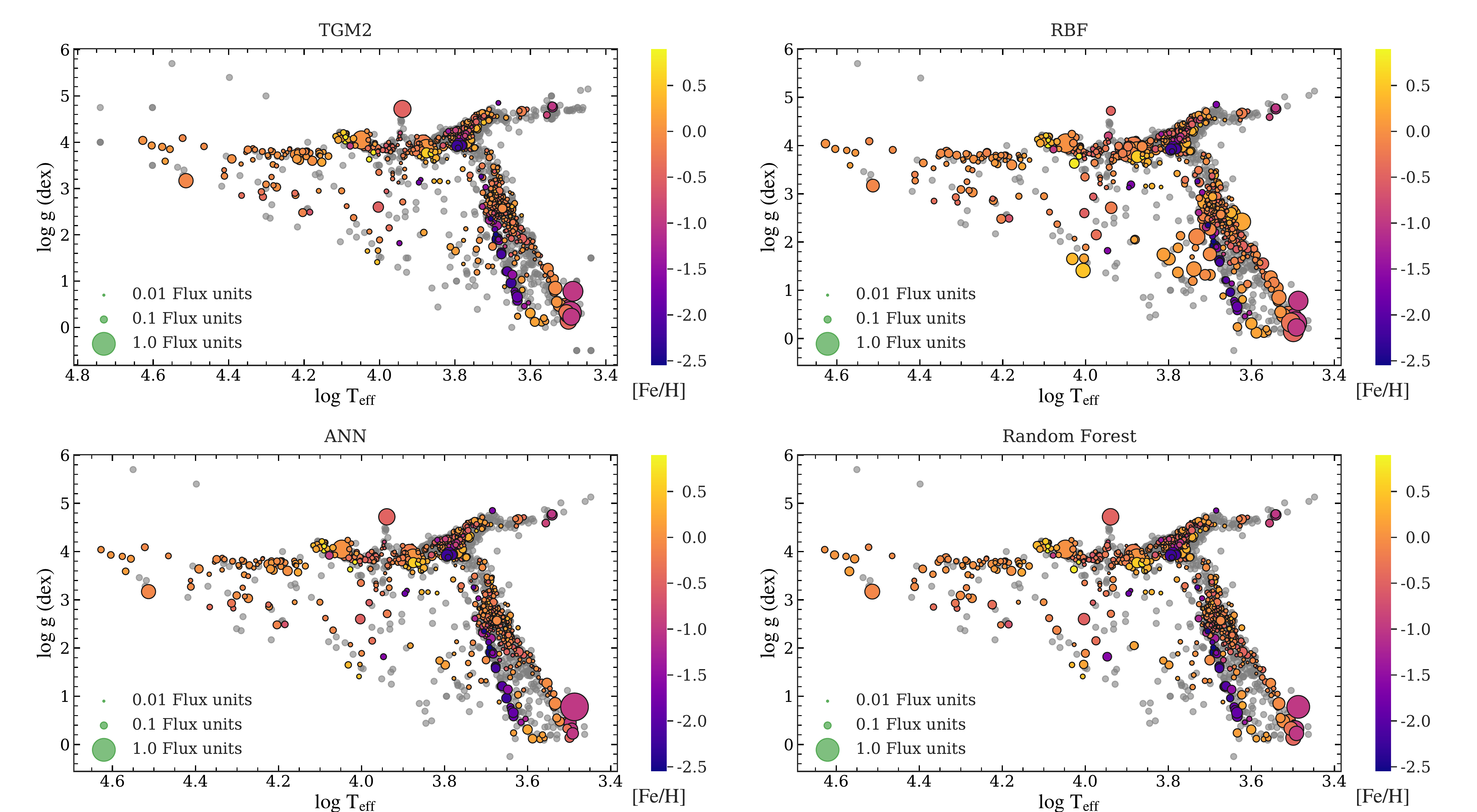}
\caption{Distribution of mean absolute errors for the test sample (CFLIB) over the parameter space using various approaches discussed in the paper: TGM2 interpolation, Gaussian RBF interpolation, ANN, and RF models. The color of the plotting symbol represents the \feh{} values of the test spectra, as shown by the color bar. Size of the plotting symbol is indicative of the mean absolute error as indicated in legend.}
  \label{fig:error_parameter_space}
\end{figure*}

To investigate the regions of parameter space where the differences are larger in the four models, we show the MAE as a function of the atmospheric parameters in Fig.~\ref{fig:error_parameter_space}. The diagram helps in understanding the most discrepant cases which remain unaccounted for while computing the robust statistics. We find that in most regions, the performance of TGM2, ANN, and RF is better than the RBF interpolation. For the coolest stars of the sample (dwarfs as well as giants), ANN's performance is marginally better than the other three methods. This region of the parameter space was the major concern with the TGM interpolator and had motivated the development of an improved version, TGM2. Overall, the ML models achieve better interpolation than the other two methods of spectral interpolation.

\subsection{Wavelength dependent systematics}\label{sec:lambda_test}

\begin{figure*}
  \centering
    \includegraphics[width=0.48\linewidth]{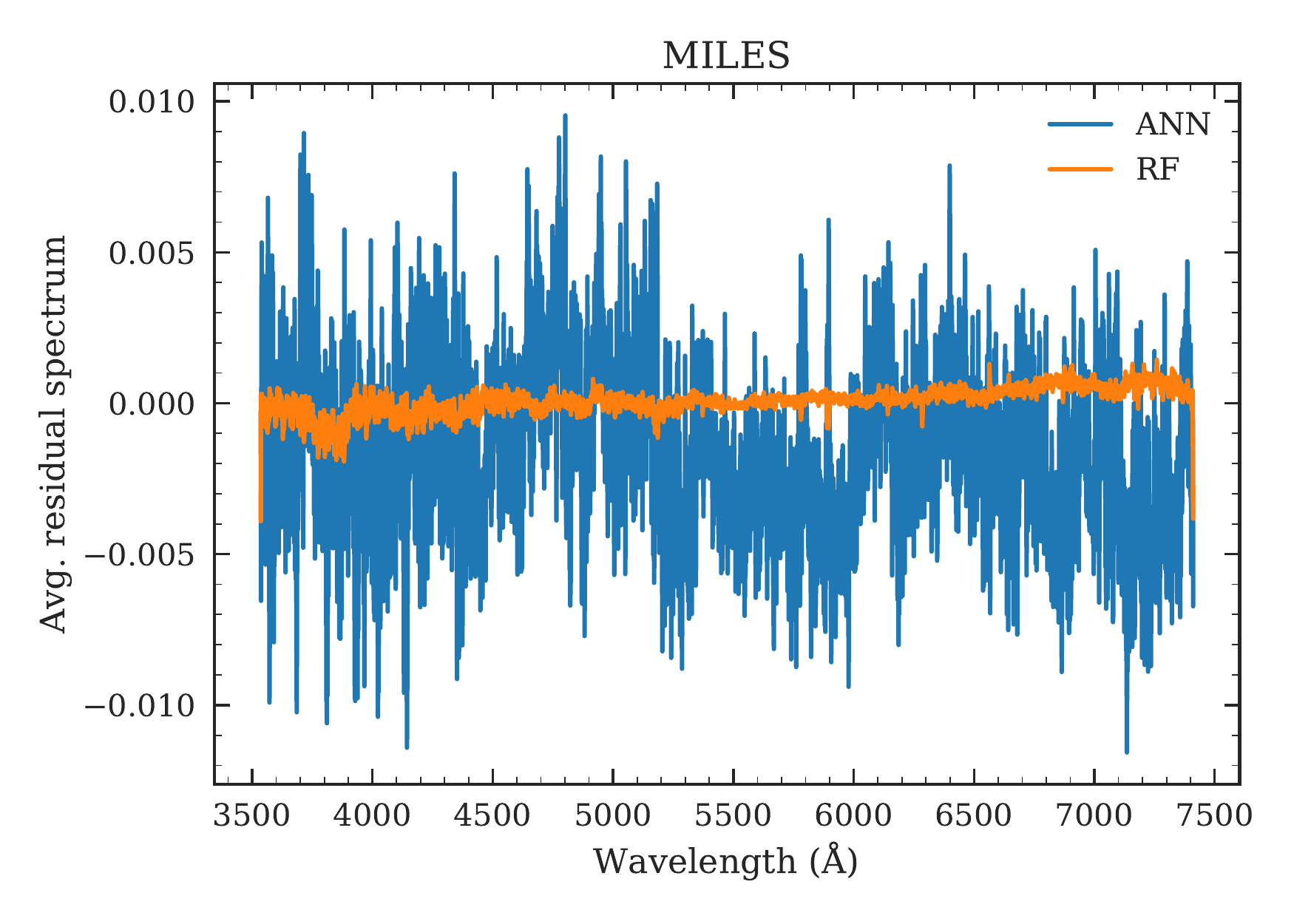}
    \includegraphics[width=0.48\linewidth]{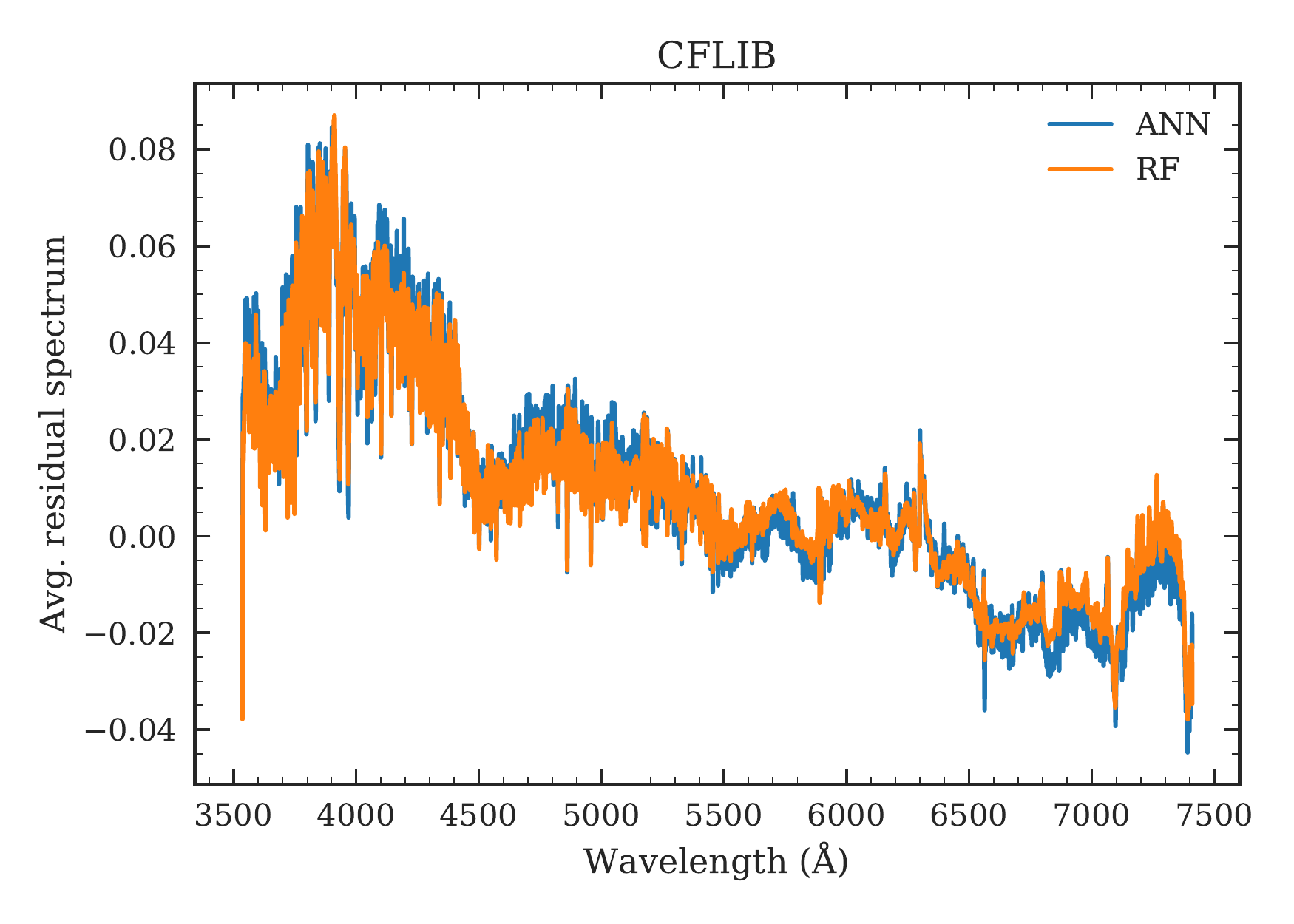}
\caption{Average residual spectrum for the training (on the left) and test set (on the right) calculated using the two ML models. ANN and RF estimated residual spectra are shown in the blue and orange color respectively as indicated in the legends.}
  \label{fig:error_lambda}
\end{figure*}

In subsection~\ref{sec:comparison}, we check the residuals between the original and interpolated spectra as a function of the three atmospheric parameters. Now, we probe the residuals as a function of the wavelength by computing the average residual spectrum using the two ML models, ANN and RF, for our training (MILES) and test set (CLFIB). The average residual spectrum is defined as the robust (outlier resistant) average of the differences between the interpolated and the original spectra (interpolated$-$original). The residual spectra are shown in Fig.~\ref{fig:error_lambda}. For the training spectra from MILES, we see that the RF performs much better than the neural network model. This is mainly due to the early stopping of the neural network training, which makes sure that the network does not overfit the training set. The learning curves for the MAE in Fig.~\ref{fig:val_curve} show that by the end of the training, the curve for the validation set flattens out, whereas the curve for the training set has still some gradient. If the training is not stopped, the NN will keep on improving on the training set without learning anything new. On the other hand, it is well known that the ensemble methods like RF can learn the training data very well with almost zero prediction variance over the training examples. This observation of RF's better performance over ANN for the training set also agrees with the metrics presented in Table~\ref{tab:inter_comp}. We notice that the residuals with either model do not show any wavelength-dependent trend. Next, we compute the average residual spectrum for the test set and find that the two residual spectra are very similar. This supports the argument that both ML models perform equally well and either of them can be used as a final model for the interpolation. However, it is evident that the differences are larger at the bluer end (<\,4500 \AA), where the flux is overestimated by about 0.08 flux units. This trend is present in both residual spectra. This systematic difference could arise due to various reasons. Three primary contributors to the discrepancy are:
\begin{enumerate}
    \item Error in the parameters/spectra of the training set,
    \item Error in the interpolation model, and
    \item Error in the input atmospheric parameters/spectra of the test set.
\end{enumerate}

We now examine each source of error in detail. As already discussed, we find only two sources of homogeneous measurements for the MILES atmospheric parameters in the literature, CEN \& PS. As shown in Table~\ref{tab:inter_comp}, the ANN/RF models trained on the PS parameters give the least residuals. Since we apply the trained model to the training sample itself in this assessment, this provides an estimate of the combined error due to (i) and (ii). The value of the RMSE using ANN on the training set is $0.0285$ in flux units, which can be considered an upper limit of the combined error (lower limit on the interpolation error on the test set). This value is smaller than the RMSE of $0.0628$ obtained for the test set. A similar difference between the training and test RMSE is seen for the RF model. Also, as seen in the left panel of Fig.~\ref{fig:error_lambda}, the average residual spectra for the training set do not show any dependence on the wavelength. The variation is uniform across all the wavelengths. Therefore we conclude that the combined error due to (i) and (ii) contribute by a meager amount to the final error on the test set. Also, the wavelength-dependent trend in the average residuals for the test set (right panel in Fig.~\ref{fig:error_lambda}) is not due to any systematic in either training set or the ML models. We would also like to emphasize that there are practical constraints on reducing the error introduced due to the error in the training set as there are only two sources of homogeneous measurements for the MILES atmospheric parameters in the literature and the error spectrum for the MILES library is not available.

\begin{figure}
  \centering
    \includegraphics[width=0.96\linewidth]{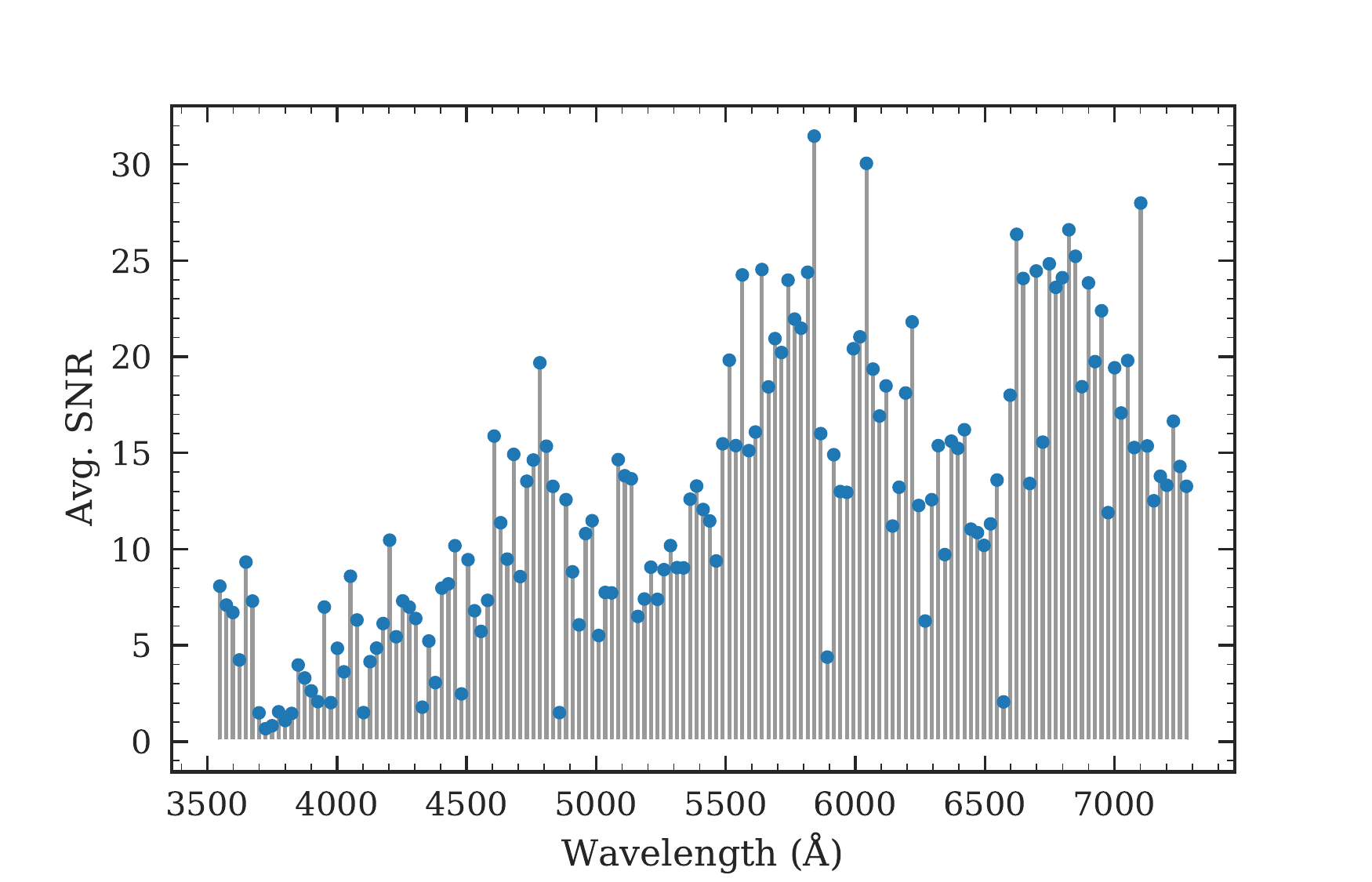}
\caption{Average signal-to-noise ratio (SNR) of test spectra from the CFLIB as a function of wavelength. SNR has been calculated in every 25 \AA{} wavelength interval.}
  \label{fig:snr}
\end{figure}

\begin{figure}
  \centering
    \includegraphics[width=0.95\linewidth]{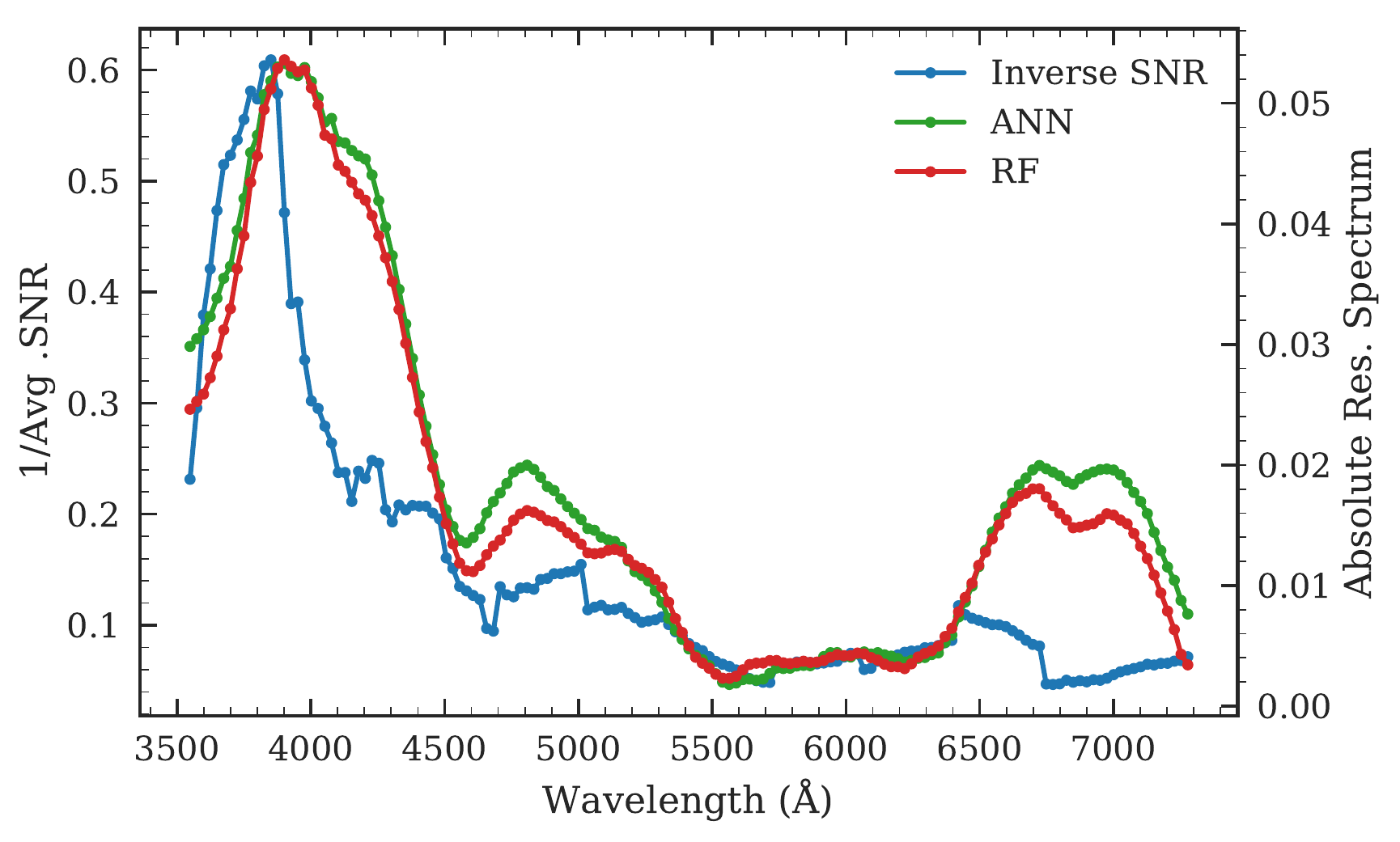}    
\caption{Smoothed variation of inverse average SNR (blue), ANN residual spectrum (green), and RF residual spectrum (red) as a function of wavelength for the CFLIB spectra. The y-axes on the left and right show the scale for inverse average SNR and the absolute residual spectra, respectively.}
  \label{fig:error_snr}
\end{figure}

Comparatively larger error on the test set is a common observation in most ML applications, but the systematically overestimated flux towards the blue end can not be attributed to any discrepancy in the training set or in the interpolation models. We also check the TGM2 and RBF interpolators to understand whether the pattern in the average residual spectrum is due to our interpolation model or is associated with the CFLIB parameters/spectra. We find that the trend with the TGM and RBF interpolators also remains the same which assures that the errors are specific to the test spectra/parameters from the CFLIB.
We further investigate this issue and postulate that the higher residuals in the blue region could be due to the lower signal-to-noise ratio (SNR). To verify it, we compute the SNR of each test spectrum in 25 \AA{} bins, which results in an SNR vector containing about 150 SNR values for the wavelength range of our study. For 850 test spectra, we get 850 SNR vectors and calculate the average SNR vector. The variation in the average SNR with respect to wavelength is shown in Fig.~\ref{fig:snr}. We notice that the average SNR is comparatively lower for wavelengths less than 4500 \AA{} and varies between 0 to 10. SNR values are maximum in the region around 5500-6200 \AA. For better visualization of error dependence on the SNR, we compute the average residual spectra (shown in Fig.~\ref{fig:error_snr}) in 25 \AA{} bins and plot their absolute values with the inverse of the average SNR vector (1/Avg. SNR Vector) in the same frame. Fig~\ref{fig:error_snr} shows the smoothed variation of the absolute average residual spectra with the inverse SNR vector. The inverse SNR vector and the residual spectra follow a similar trend, which tallies with our expectations that the poor SNR in the blue region manifests itself in the form of higher residuals and vice-versa.
We also note that five grating settings were used while observing the CFLIB spectra to cover the wavelength range around 3400\,–\,9500 \AA{} \citep{Valdes2004}. In some cases, there was no overlap between the adjacent grating settings, which resulted in small gaps of about 50 \AA. We conclude that the trends in the residual spectra can be explained by considering the varying SNR across wavelengths along with the flux-calibration issues (Sec.~\ref{sec:outliers}) and gaps in the CFLIB spectra.

\subsection{Model sensitivity to the error in the atmospheric parameters}

For estimating the stellar atmospheric parameters, spectroscopically measured atmospheric parameters are considered to be more reliable than the photometry-based measurements \citep{Sharma2016,Balona2016,Joshi2019}. In case the measurements from spectroscopy are not available, the photometrically derived parameters are adopted. Independent of the method used for the determination, these parameters always come with some uncertainty in their values. Since these parameters are used as an input to the interpolation model, the error in the parameters also propagates to the interpolated spectrum. It is essential for an interpolation model to be insensitive towards the small errors in the input parameters, which in turn shows the robustness of the model for a wider range of applications.

\begin{figure*}
  \centering
    \includegraphics[width=0.48\linewidth]{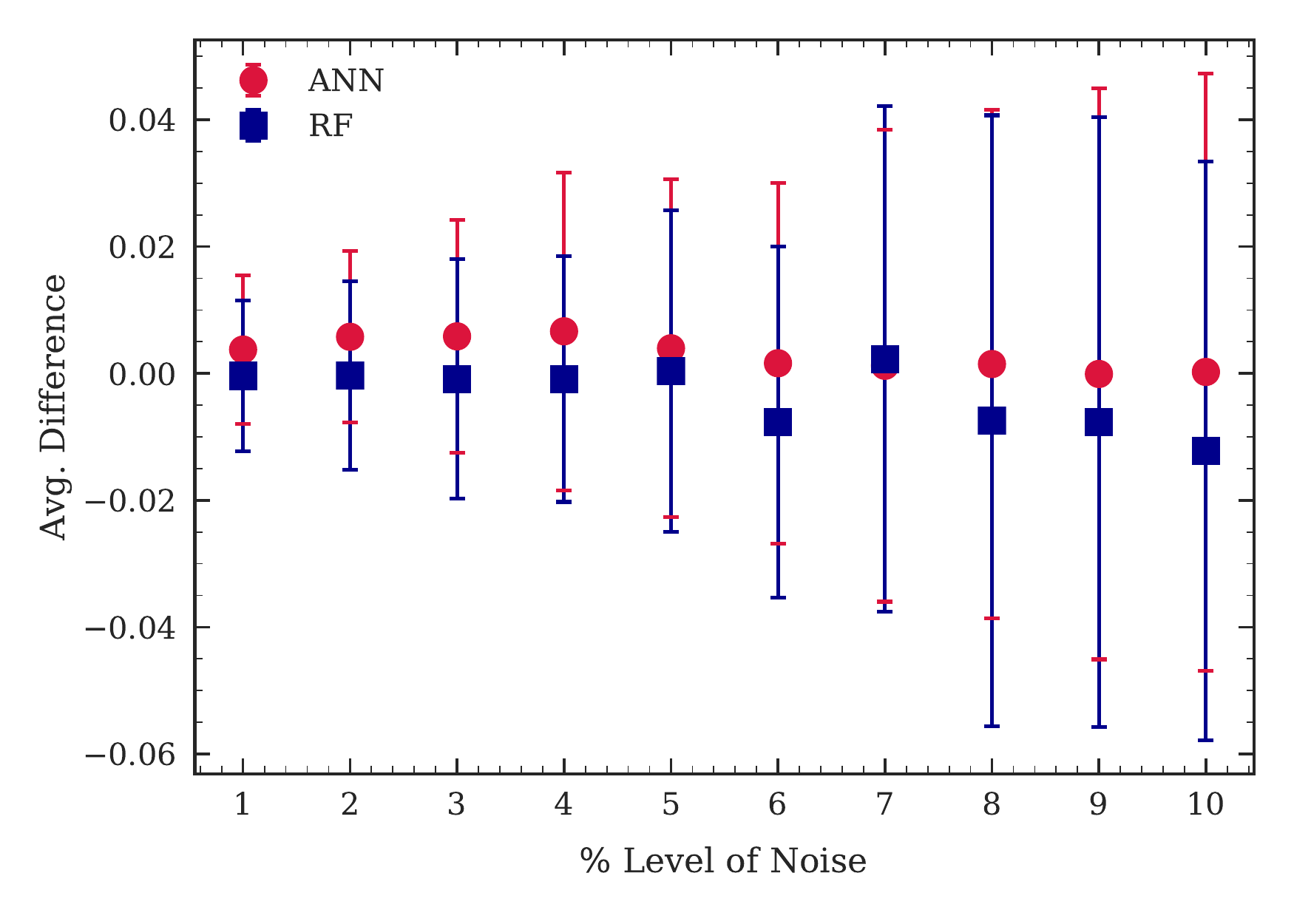}
    \includegraphics[width=0.48\linewidth]{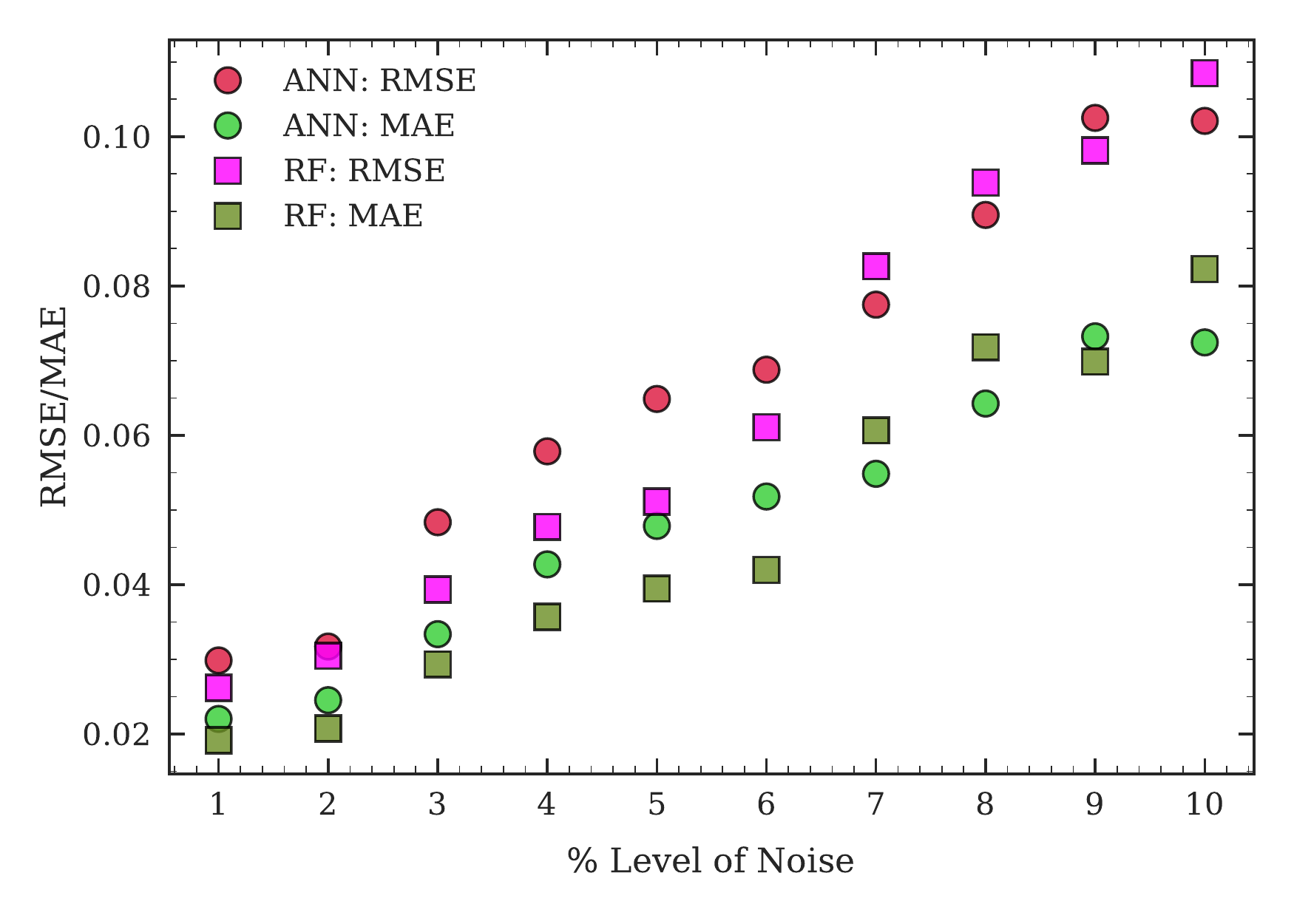}
\caption{Errors in the interpolated spectra using NN (circles) and RF (squares) models with different noise levels in the input atmospheric parameters. Left panel shows the variation in the average difference (interpolated$-$original) as a function of noise in the input atmospheric parameters and the error bars indicate the standard deviation. The RMSE and MAE values are presented in the right panel.}
  \label{fig:noise}
\end{figure*}

To evaluate the robustness of the ML models against uncertainty in the input parameters, we use the spectra from the MILES library for training and testing. We randomly select 90\% of the library spectra for training and the remaining 10\% spectra for the testing. While splitting the library spectra into training and test set, we make sure that the coverage over the parameter space is the same for both sets. The choice of using the training and test spectra from the same library is primarily driven by the fact that this approach will help in analysing the errors in the interpolated spectrum solely due to the uncertainty in the input atmospheric parameters and will have the least effect due to other factors like different instruments, observational errors, data reduction strategies, etc. We use the same ANN architecture as prescribed in Sec~\ref{sec:ann_keras} and retrain it on the new training set with 885 spectra. We introduce Gaussian noise in the three atmospheric parameters of the test set spectra at ten different levels varying from 1\% to 10\%. An error of 1\% means that the values of three parameters might be uncertain within 1\% of their actual values. A spectrum with \teff{}\,=\,5000\,K will have an uncertainty up to 50\,K. Similarly, 10\% error in actual \teff{} of 18000\,K would mean that the perturbed \teff{} is within 18000$\pm$1800\,K. The same explanation holds true for the other two parameters, \logg{} and \feh{}. It is to be noted here that the noise is introduced simultaneously in the three parameters. We use MAE, RMSE, and average of the difference between the interpolated and original test spectra as metrics to study the effect of introducing the noise in the input parameters. The variation in these metrics at different levels of noise is shown in Fig.~\ref{fig:noise}. We find that the value of average difference remains almost at the zero level but the dispersion increases from about 0.02 to 0.05 flux units, which is not surprising as the input parameters have been perturbed from their actual values. The value of MAE also increases from about 0.02 to 0.07 for the introduced noise in the range of 1\% to 10\%. MAE seems to follow a linear trend indicating a linear increase in the interpolation error with increasing noise. A similar trend is observed for the RMSE as well. We repeat the same analysis using the RF model as well and find no significant difference qualitatively as well as quantitatively in the model behaviour to the introduced errors.

This test provides an upper bound for the errors in the interpolated spectra purely due to the noise in the atmospheric parameters which could be useful in disentangling the contribution from the uncertainties in the parameters and issues with the observed spectra. The analysis shows that the ML models presented here are quite robust to small errors in the input atmospheric parameters and their behaviour does not change dramatically with the introduction of noise.

\section{Conclusions and Discussion}\label{sec:conclusion_discussion}

We have trained a fully connected neural network model and a random forest model for generating a stellar spectrum for a given set of three atmospheric parameters: \teff, \logg, and \feh. We use the MILES spectral library for training the network, with input atmospheric parameters taken from \citet{Prugniel2011} and \citet{Sharma2016}. For an independent examination of the model validity, we use the Indo-U.S. Library of Coud{\'e} Feed Stellar Spectra. Homogeneous parameter determination for the CFLIB spectra has been provided in \citet{Wu2011}. Using these parameters with the trained models, we generate CFLIB spectra and compare them with the observed spectra. We choose RMSE, MAE, Avg. difference between the two sets of spectra and $R^2$ score for assessing the performance of the models and their inter-comparison. Based on the comparison between different interpolation models, namely ANN, RF, TGM2, and RBF, we find that the ANN and RF models presented in this work perform significantly better than TGM2 and RBF interpolation methods. The trained ANN model gives an average difference of $-0.0126\pm 0.0268$ in the normalized flux units on the test set with mean absolute error of 0.0395. The optimized RF model tested on the CFLIB spectra returns an average difference of $-0.0123\pm0.0261$  between the original and generated spectra with mean absolute error of 0.0392. The performance of the two models is comparable and the error statistics indicate an overall good quality reconstruction. With both the models, We obtain $R^2$ score of about 0.96, which means that 96\% of the variance in the original spectra can be accounted by the model spectra, implying very good agreement between the two.

We also study the most discrepant cases (with the largest residuals) from the test set and find that only $\sim$2\% of the spectra in the whole test set are not reconstructed satisfactorily. The inconsistency primarily occurs only for the M giants which is the region in the parameter space populated by only six examples in our training data. The sparse density of these type of spectra in the training set makes it difficult for the model to generate the spectrum as efficiently as in the other regions of the space. We also show that the large deviations with respect to the CFLIB spectra shown in some cases with \teff{}\,>\,7500\,K are the result of improper flux calibration in the original spectra.
We arrive at the same conclusion when we study the mean absolute interpolation error as a function of three stellar parameters. The regions occupied by the cool giants and hot dwarfs are sparsely populated, which results in an overall increase in the interpolation error in these regions. We also examine the interpolation error as a function of wavelength by computing the average residual spectrum. For the training set, we find that the residual spectra for both the models show a uniform variation across all wavelengths. For the test set, we observe that, on average, the flux values near the blue wavelengths (< 4500 \AA) are getting overestimated, whereas the flux in the red region is marginally underestimated. We trace the origin of this trend to the fluctuation in the signal-to-noise ratio, which is relatively lower in the blue region in our test spectra. Varying SNR combined with other factors like the sparse density in certain regions and improper flux-calibration gives rise to higher interpolation error for the test spectra as compared to the training set spectra.
Other than these primary causes, various other components such as gaps in the test spectra from the CFLIB, uncertainties in the atmospheric parameters, internal dispersion between the parameters estimated using different techniques, extinction correction, specificity of a star, effects in a particular spectrum, etc. also contribute to the residuals.
To estimate how much interpolation error can enter in our models due to uncertainties in the input parameters, we train and test on the spectra from the same library, MILES in our case, and find that the MAE can go from 0.02 to 0.07 flux units with 1-10\% error in the input parameters. Previous studies by \citet{Prugniel2011} and \citet{Wu2011} that use the full-spectrum fitting technique to obtain the stellar parameters exhibit that the errors from the low-resolution spectroscopic studies can vary from <1\% to 3\%. Within these limits, the performance of our models is reasonably well. In this study, we use the empirical spectra for the reasons already discussed in Sec.~\ref{sec:introduction}. Our choice of the MILES library as the training set is guided by the fact that the spectral resolution of MILES is similar to the large spectroscopic surveys like the SDSS and the LAMOST which are the prime target for such automated frameworks. However, we also train and test the same NN architecture presented in Sec.~\ref{sec:ann_keras} on theoretical spectra to compare it with the previous studies. These spectra were part of the Sloan Extension for Galactic Exploration and Understanding (SEGUE) Stellar Parameter Pipeline \citep[SSSP;][]{Lee2008} and were used to estimate the atmospheric parameters for the SDSS spectra. Training NN on 90\% (724) spectra and testing on the remaining 10\% (81) of the spectra reveals rms of 0.4\%. This is comparable to the rms error of 0.1\% by \textit{The Payne} reported in \citet{Ting2019}. It is important to note here that \textit{The Payne} uses 25 stellar labels to interpolate the spectra while in our test, we only used three atmospheric parameters. We believe that the absence of abundances (currently unavailable) is the major reason for the difference in the error between the two studies and including these can bring down the error of 0.4\% significantly. 

There are three significant improvements that can be considered in a future study: (i) Enhancing the performance of developed ML models in the low-density regions which can be achieved by introducing a weighting scheme to assign larger weights to the input parameters with lesser number of training examples. The weights could be automatically assigned using some technique, like Gaussian-kernel smoothing, which can effectively measure the density in any location of the parameter space; (ii) Adding extrapolation support for the cooler stars by incorporating spectra from external archives; (iii) \citet{Ting2019} have shown that using other stellar parameters and individual elemental abundances can significantly lower the interpolation error which translates to a lesser error in the parameter determination. Though with the low-resolution empirical spectra, it is difficult to estimate the elemental abundances but studies like \citet{Milone2011, Ting2017b, Ting2017a}, which obtained individual elemental abundances from low-resolution spectra, can potentially be used to refine the current models.

The trained ML models can be advantageous in various stellar spectral analyses. One important application of this model would be to determine the atmospheric parameters of observed stellar spectra by comparing them with template spectra prepared using a model for different atmospheric parameters. The traditional approach of spectral interpolation requires fine-tuning of various coefficients and/or terms of the polynomial if one wants to incorporate another spectral library for the interpolation grid. But because of the scalability and generalization capability of the algorithms we have used, the process of integrating stellar spectra from different libraries and retraining the model becomes easier. Moreover, the computation time for training the ML models and generating the spectra from the trained models is significantly shorter than the polynomial based interpolator. On the same machine, the ML models take about 2-3 seconds for producing the spectra of 984 MILES stars, whereas TGM2 takes 24.6 seconds (0.025 seconds for each spectrum on average). This aspect makes ML-based models more favorable for the large spectroscopic databases, like LAMOST and Sloan Digital Sky Survey \citep[SDSS,][]{York2000}, which employ their own data reduction pipelines for delivering the primary science results from the observed spectra. Another application of these models would be to the recalibration of the observed spectra with improper flux-calibration or gaps in spectral coverage as highlighted in this work. This can enhance the quality of the existing stellar spectral libraries and, in turn, might lead to improved results from various studies involving these libraries.

\section*{Acknowledgements}
We thank Philippe Prugniel (Observatoire de Lyon, Lyon, France), Ariane Lan\c{c}on (Observatoire Astronomique Strasbourg, Strasbourg, France), and T. Sivarani (Indian Institute of Astrophysics, Bengaluru, India) for a critical review of the manuscript and providing important inputs. We are thankful to T. Sivarani for providing the theoretical spectra used in SSPP and Liantao Cheng (Yunnan Observatories, Kunming, China) for giving us the access to their RBF interpolation code. KS and AK acknowledge financial support from a Raja Ramanna Fellowship (10/1(16)/2016/RRF-R\&D-II/630) awarded by the Department of Atomic Energy, Government of India. HPS and RG thank the National Astronomical Observatories, Chinese Academy of Sciences, Beijing for special guest researcher grants and for kind hospitality. HPS thanks CSIR for support through grant 03(1428)/18/EMR-II. KV was affiliated with IUCAA during the bulk of the work under a position funded by the National Knowledge Network's data driven initiatives in Astronomy and Biology grant.
We thank the anonymous referee for the careful review and suggestions that
significantly improved the quality of this work.

This work has made use of the LAMOST database. The Guoshoujing Telescope (the Large Sky Area Multi-Object Fiber Spectroscopic Telescope; LAMOST) is a National Major Scientific Project built by the Chinese Academy of Sciences. Funding for the project has been provided by the National Development and Reform Commission. LAMOST is operated and managed by the National Astronomical Observatories, Chinese Academy of Sciences. 

\section*{Data Availability}

The models presented in this article can be accessed via the URL: \url{https://ddi.iucaa.in/spectralInterpolator}.

\bibliographystyle{mnras}
\bibliography{bibliography}


\label{lastpage}

\end{document}